\documentclass[preprint,3p,12pt]{elsarticle}
\usepackage{amsmath}
\usepackage{stmaryrd}
\usepackage{bbding}
\usepackage{dcolumn}
\usepackage{graphicx}
\usepackage{amsfonts}
\usepackage{amssymb}
\usepackage{psfrag}
\usepackage{wrapfig}
\usepackage{subfigure}
\usepackage{makeidx}
\usepackage{bm}
\usepackage{epsf}
\usepackage{epsfig}
\usepackage{setspace}
\usepackage{color}
\usepackage{epstopdf}
\usepackage{lineno}

\begin{document}
\title{Modeling and simulation in supersonic three-temperature carbon dioxide turbulent channel flow}

\author[SUSTech1,SUSTech2]{Guiyu Cao}
\ead{caogy@sustech.edu.cn}
\author[PKU]{Yipeng Shi\corref{cor}}
\ead{syp@mech.pku.edu.cn}
\author[HKUST]{Kun Xu}
\ead{makxu@ust.hk}
\author[SUSTech1,SUSTech2]{Shiyi Chen\corref{cor}}
\ead{chensy@sustc.edu.cn}

\address[SUSTech1]{Academy for Advanced Interdisciplinary Studies, Southern University of Science and Technology, Shenzhen, Guangdong 518055, People’s Republic of China}
\address[SUSTech2]{Guangdong-Hong Kong-Macao Joint Laboratory for Data-Driven Fluid Mechanics and Engineering Applications, Southern University of Science and Technology, Shenzhen, Guangdong 518055, People’s Republic of China}
\address[PKU]{Department of Aeronautics \& Astronautics Engineering, Peking University, Beijing 100871, People’s Republic of China}
\address[HKUST]{Department of Mathematics, Hong Kong University of Science and Technology, Clear Water Bay, Kowloon, Hong Kong}
\cortext[cor]{Corresponding author}

\begin{abstract}
This paper pioneers the direct numerical simulation (DNS) and physical analysis in supersonic three-temperature carbon dioxide ($CO_2$) turbulent channel flow.
$CO_2$ is a linear and symmetric triatomic molecular, with the thermal non-equilibrium three-temperature effects arising from the interactions among translational, rotational and vibrational modes under room temperature.
Thus, the rotational and vibrational modes of $CO_2$ are addressed.
Thermal non-equilibrium effect of $CO_2$ has been modeled in an extended three-temperature BGK-type model, with the calibrated translational, rotational and vibrational relaxation time. 
To solve the extended BGK-type equation accurately and robustly, non-equilibrium high-accuracy gas-kinetic scheme is proposed within the well-established two-stage fourth-order framework.
Compared with the one-temperature supersonic turbulent channel flow, supersonic three-temperature $CO_2$ turbulence enlarges the ensemble heat transfer of the wall by approximate $20\%$, and slightly decreases the ensemble frictional force.
The ensemble density and temperature fields are greatly affected, and there is little change in Van Driest transformation of streamwise velocity.
The thermal non-equilibrium three-temperature effects of $CO_2$ also suppress the peak of normalized root-mean-square of density and temperature, normalized turbulent intensities and Reynolds stress.
The vibrational modes of $CO_2$ behave quite differently with rotational and translational modes.
Compared with the vibrational temperature fields, the rotational temperature fields have the higher similarity with translational temperature fields, especially in temperature amplitude.
Current thermal non-equilibrium models, high-accuracy DNS and physical analysis in supersonic $CO_2$ turbulent flow can act as the benchmark for the long-term applicability of compressible $CO_2$ turbulence.
\end{abstract}

\maketitle

\section{Introduction}
Mars exploration programs are currently experiencing a revival, such as amazing Mars robotic helicopter "Ingenuity" operating on Mars \cite{Ingenuity}.
Mars's atmosphere consists of $95.32\%$ carbon dioxide ($CO_2$).
For accurate predictions of surface drag and heat flux on Martian vehicles, it is necessary to take the peculiarities of $CO_2$ into account.
Different with the dominant diatomic gases nitrogen ($N_2$) and oxygen ($O_2$) on earth, $CO_2$ is a linear and symmetric triatomic molecular, which has three vibrational modes \cite{atkins2006physical}.
Triatomic molecular $CO_2$ is equipped with the inherent thermal non-equilibrium multi-temperature effects arising from the interactions among the translational, rotational and vibrational modes \cite{kustova2006correct, kustova2019relaxation}.

Carbon dioxide is widely studied in the applications of physical chemistry and fluid dynamics, i.e., $CO_2$-$N_2$ gas laser system \cite{taylor1969survey}, environmental green-house problems \cite{stouffer1989interhemispheric}, and Mars entry vehicles \cite{candler1990computation, chen1995mars, braun2006mars}.
Complex molecular structure and multiple internal energy relaxation mechanisms in $CO2$ significantly affect its physical and chemical properties \cite{kustova2019models}.
In fluids community, of special interest is the evaluation of bulk viscosity in $CO2$.
Stokes’ viscosity relation does not hold for $CO_2$ \cite{tisza1942supersonic}, as the bulk viscosity can be thousands of times larger than the shear viscosity. 
Many research works are engaged in experimental and theoretical studies in the bulk viscosity of $CO_2$ \cite{pan2005power, cramer2012numerical}.
Recent experiments \cite{jamali2019rayleigh, wang2019bulk} show that the vibrational modes contribute dominantly to the bulk viscosity of $CO_2$, and the bulk viscosity from rotational modes is only one half of its shear viscosity approximately \cite{wang2019bulk}.
Thus, the vibrational modes should be modelled carefully when simulating $CO_2$ flows.
In view of the importance of multiple internal energy modes, the multiple internal energy relaxation mechanisms of $CO_2$ \cite{kustova2006correct, kustova2019relaxation} have been modeled and analyzed, which confirm that equilibrium one-temperature gas flow description is not valid for $CO_2$ flows even under room temperature (i.e., $300K$).
To the author's knowledge, the thermal non-equilibrium physical models considering the multi-temperature effects of $CO_2$ and its applications in turbulent flows are seldom reported.
For accurate predictions of $CO_2$ turbulence, it is necessary to take the  three-temperature effects of $CO_2$ into account.

In the past few decades, the gas-kinetic scheme (GKS) based on the Bhatnagar-Gross-Krook (BGK) model \cite{bhatnagar1954model, chapman1970mathematical} 
has been developed systematically for the computations from low speed flows to hypersonic ones \cite{xu2001gas, xu2015direct}.
Based on the time-dependent flux solver, including generalized Riemann problem solver and GKS \cite{li2016two, pan2016efficient}, a reliable two-stage fourth-order framework was provided for developing the high-order GKS (HGKS) into fourth-order accuracy.
With the advantage of finite volume GKS and HGKS, they have been naturally implemented as a direct numerical simulation (DNS) tool in simulating turbulent flows \cite{fu2006numerical, liao2009gas, kumar2013weno}, especially for compressible turbulence \cite{cao2019three, cao2021three}.
Aiming to conduct the large-scale DNS, a parallel in-house computational platform of HGKS has been developed in uniform grids and curvilinear grids \cite{cao2021high, cao2022high}, with high efficiency, fourth-order accuracy and super robustness.
In addition, with the discretization of particle velocity space, a unified gas-kinetic scheme (UGKS) \cite{xu2010unified, huang2013unified} and unified gas-kinetic wave particle method (UGKWP) \cite{zhu2019unified, liu2020unified, yang2022unified} have been developed for multi-scale physical transport problems.
The well-developed HGKS and multi-scale UGKS/UGKWP provides the solid foundation for thermal non-equilibrium multi-temperature modeling and simulation in $CO_2$ flows.
The multi-scale modeling and numerical framework can be applied in multi-scale $CO_2$ flows, i.e., the Mar's re-entry vehicles from rarefied to continuum regimes.
As a starter, current study focuses on the supersonic $CO_2$ turbulence in the continuum regime.

In this paper, the vibrational modes of $CO_2$ are addressed, and the translational, rotational and vibrational relaxation time of $CO_2$ are calibrated.
The three-temperature effects of $CO_2$ are modeled in an extended three-temperature BGK-type model within the well-established kinetic framework \cite{cai2008one, cao2018physical, liu2021multiple}.
To achieve high-order accuracy in space and time for simulating the supersonic $CO_2$ turbulence, the non-equilibrium high-accuracy GKS has been constructed with the second-order kinetic flux, fifth-order WENO-Z reconstruction \cite{castro2011high}, and two-stage fourth-order time discretization \cite{pan2016efficient}.
One-temperature supersonic turbulent channel flow \cite{coleman1995numerical, zhang2020contribution} is simulated firstly to validate the numerical set-up with bulk Mach number $Ma = 3$ and bulk Reynolds number $Re = 4880$.
Considering the translational, rotational, and vibrational specific heats at constant volume, one-temperature supersonic turbulent channel flow
of thermally perfect gas has been studied \cite{chen2019effects}. 
With implementing the non-equilibrium high-accuracy GKS in the large-scale parallel in-house platform \cite{cao2021high, cao2022high}, for the first time, the DNS in supersonic three-temperature $CO_2$ turbulent channel flow is conducted.
Compared with the one-temperature supersonic turbulent channel flow, the three-temperature effects of $CO_2$ are analyzed.
Numerical simulation confirms the thermal non-equilibrium three-temperature performance of $CO_2$.
Both the maximum ensemble temperature and normalized r.m.s. temperature sort from high to low is translational temperature, rotational temperature, and vibrational temperature. 
Compared with the vibrational temperature fields, the rotational temperature fields has the higher similarity with translational temperature fields both in temperature amplitude and its structure.

For physical modeling and numerical simulation in supersonic three-temperature $CO_2$ turbulent channel flow, this paper is organized as follows. 
Section 2 addresses the internal energy modes of $CO_2$. 
Extended thermal non-equilibrium three-temperature BGK-type model and corresponding  non-equilibrium high-accuracy GKS for $CO_2$ are included in Section 3. 
Numerical examples and discussions are presented in Section 4. 
The last section is the conclusion and remarks.

\section{Internal energy modes of carbon dioxide}
Thermal non-equilibrium three-temperature effects of $CO_2$ mainly arise from the interactions among internal energy modes \cite{kustova2019relaxation}.
This section addresses the vibrational modes of $CO_2$, and focuses on the translational, rotational and vibrational relaxation time for the extended three-temperature BGK-type model.

\subsection{Rotational and vibrational modes}\label{subsection_rvm}
Carbon dioxide is a linear and symmetric triatomic molecular, with rotational and vibrational internal energy modes.
The characteristic rotational temperature is defined as $\theta_{r} = h_P^2 /(8\pi^2 k_B I)$, where $h_P$ is the Planck constant, $k_B$ the Boltzmann constant, $I$ the molecular moment of inertia.
For carbon dioxide, $\theta_{r} = 0.56K$ can be obtained \cite{atkins2006physical}, while the $\theta_{r}$ for $N_2$ and $O_2$ is $2.88K$ and $2.08K$, respectively.
Under room temperature, it is well known that the rotational degrees of freedom (d.o.f.) are assumed to be excited completely for $N_2$ and $O_2$.
Since $CO_2$ is with the smaller characteristic rotational temperature, the rotational d.o.f. of $CO_2$ are regarded as complete excitation in current study (i.e., $CO_2$ gas temperature above $300K$).

Carbon dioxide is equipped with three vibrational modes as one symmetric stretching mode $\nu_1$, one double degenerated bending mode $\nu_2$, and one asymmetric stretching mode $\nu_3$.
The characteristic vibrational temperature reads $\theta_{v} = h_P \tilde{\nu} c_L/k_B$, where $\tilde{\nu}$ is the characteristic wavenumber, and $c_L$ the speed of light in the vacuum.
In experimental studies of $CO_2$, infrared spectrum gives the wavelength for corresponding vibrational modes $\nu_2$ and $\nu_3$, and Raman spectroscopy provides the wavelength for $\nu_1$ \cite{atkins2006physical}.
Table \ref{tab_co2_vibrational_mode} presents the characteristic wavenumber, wavelength ($\lambda = 1/\tilde{\nu}$),  characteristic vibrational temperature and corresponding degeneracy of $CO_2$.
The double degenerated bending modes $\nu_2$ are most likely to be activated with characteristic vibrational temperature $\theta_{v} = 959.66K$, which is much lower than that of $N_2$ with $\theta_{v} = 3521K$ and $O_2$ with $\theta_{v} = 2256K$.
Thus, the vibrational modes of $N_2$ and $O_2$ is usually considered in high-temperature applications, i.e., re-entry vehicles experiencing the temperature above $800K$ \cite{anderson2000hypersonic}.
However, the excitation of vibrational modes of $CO_2$ requires to be modeled and simulated even under the room temperature \cite{kustova2019relaxation, wang2019bulk}.
\begin{table}[!h]
	\centering
	\begin{tabular}{c|c|c|c|c}
		\hline \hline
		Vibrational mode         &Wavenumber($\tilde{\nu}$)/$cm^{-1}$   &Wavelength($\lambda$)/$\mu m$    &$\theta_{v}$/$K$   &Degeneracy    \\
		\hline
		$\nu_1$      &$1388$                 &$7.20$                  &$1997.02$                           &$1$    \\
		\hline
		$\nu_2$      &$667$                  &$14.99$                   &$959.66$                            &$2$   \\
		\hline
		$\nu_3$      &$2349$                  &$4.26$                  &$3379.69$                           &$1$     \\
		\hline \hline
	\end{tabular}
	\caption{\label{tab_co2_vibrational_mode} Parameters for three vibrational modes of $CO_2$.}
\end{table}

With the assumption that there is a unique vibrational temperature at each point in the flow fields, the translational internal energy $E_t$, rotational internal energy $E_r$, and vibrational internal energy $E_v$ per unit mass of $CO_2$ read
\begin{align}
	E_t &= \frac{N_t}{2} R T_t, \label{Etrv_co2_t} \\
	E_r &= \frac{N_r}{2} R T_r, \label{Etrv_co2_t} \\
	E_v &= R \sum_{i = 1}^{3} g_i \frac{\theta_{v, i}}{e^{\theta_{v, i}/T_v} - 1}, \label{Etrv_co2_v}
\end{align}
with translational d.o.f. as $N_t$, rotational d.o.f. as $N_r$, and vibrational d.o.f. as $N_v$, where $T_t$, $T_r$ and $T_v$ represent translational temperature, rotational temperature and vibrational temperature, respectively.
In Eq.\eqref{Etrv_co2_v}, $\theta_{v, i}$ is the characteristic vibrational temperature of vibrational mode $\nu_i$, and $g_i$ is the corresponding degeneracy of mode $\nu_i$ as shown in Table \ref{tab_co2_vibrational_mode}.
The total specific heat at constant volume $C_V$ can be obtained by the sum of three components as
\begin{align} \label{Cv_co2_total}
	C_V = C_{V, t} + C_{V, r} + C_{V, v},
\end{align}
with
	\begin{align}
		C_{V, t} &= \frac{N_t}{2} R, \label{Cv_co2_t} \\
		C_{V, r} &= \frac{N_r}{2} R, \label{Cv_co2_r} \\
		C_{V, v} &= \sum_{i = 1}^{3} g_i \frac{(\theta_{v, i}/T_v)^2}{e^{\theta_{v, i}/T_v} + e^{-\theta_{v, i}/T_v} - 2} R, \label{Cv_co2_v}
	\end{align}
where $C_{V, t}$, $C_{V, r}$ and $C_{V, v}$ denotes the componential specific heat at constant volume for translational internal energy, rotational internal energy and vibrational internal energy, respectively.
In current study, the supersonic $CO_2$ turbulence is considered above $300K$, thus, translational d.o.f. as $N_t = 3$ and rotational d.o.f. as $N_r = 2$ are adopted.
Vibrational d.o.f. as $N_v$ can be obtained by the definition as $C_{V, v} = N_v R/2$, and the specific heat ratio $\gamma$ is the function of vibrational d.o.f. as
	\begin{align}
		N_v &= 2 \sum_{i = 1}^{3} g_i \frac{(\theta_{v, i}/T_v)^2}{e^{\theta_{v, i}/T_v} + e^{-\theta_{v, i}/T_v} - 2}, \label{Nv_co2} \\
		\gamma &= \frac{C_P}{C_V} = 1 + \frac{2}{5 + N_v}. \label{gamma_co2}		
	\end{align}
In Eq.\eqref{gamma_co2}, $C_P$ is the total specific heat at constant pressure.
In the high-temperature limit, note that the vibrational d.o.f. as $N_v$ approaches to classical definition with $2 \sum_{i = 1}^{3} g_i \frac{\theta_{v, i}/T_v}{e^{\theta_{vib, i}/T_v} - 1}$.

\begin{figure}[!h]
	\centering
	\includegraphics[width=0.485\textwidth]{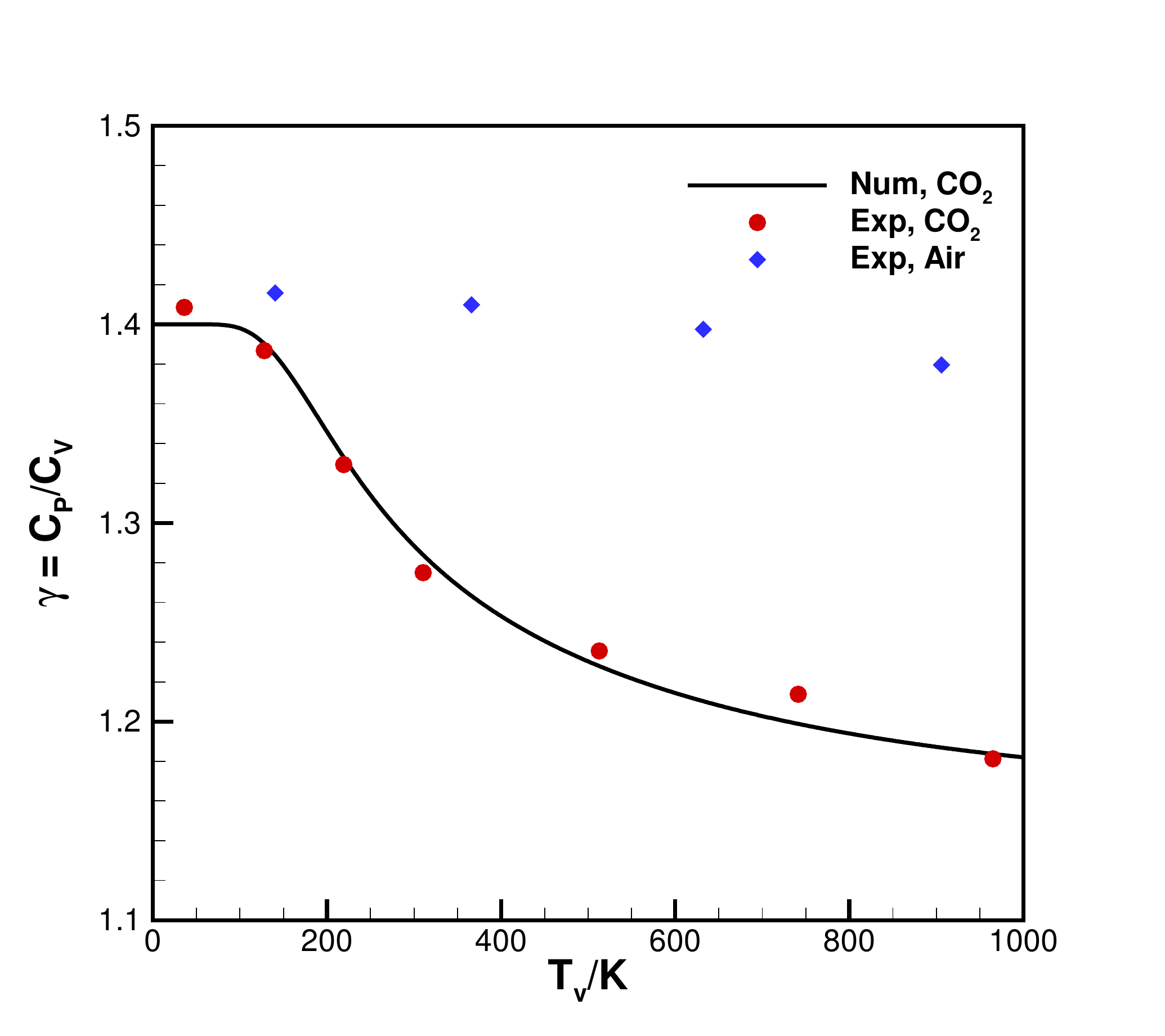}
	\includegraphics[width=0.485\textwidth]{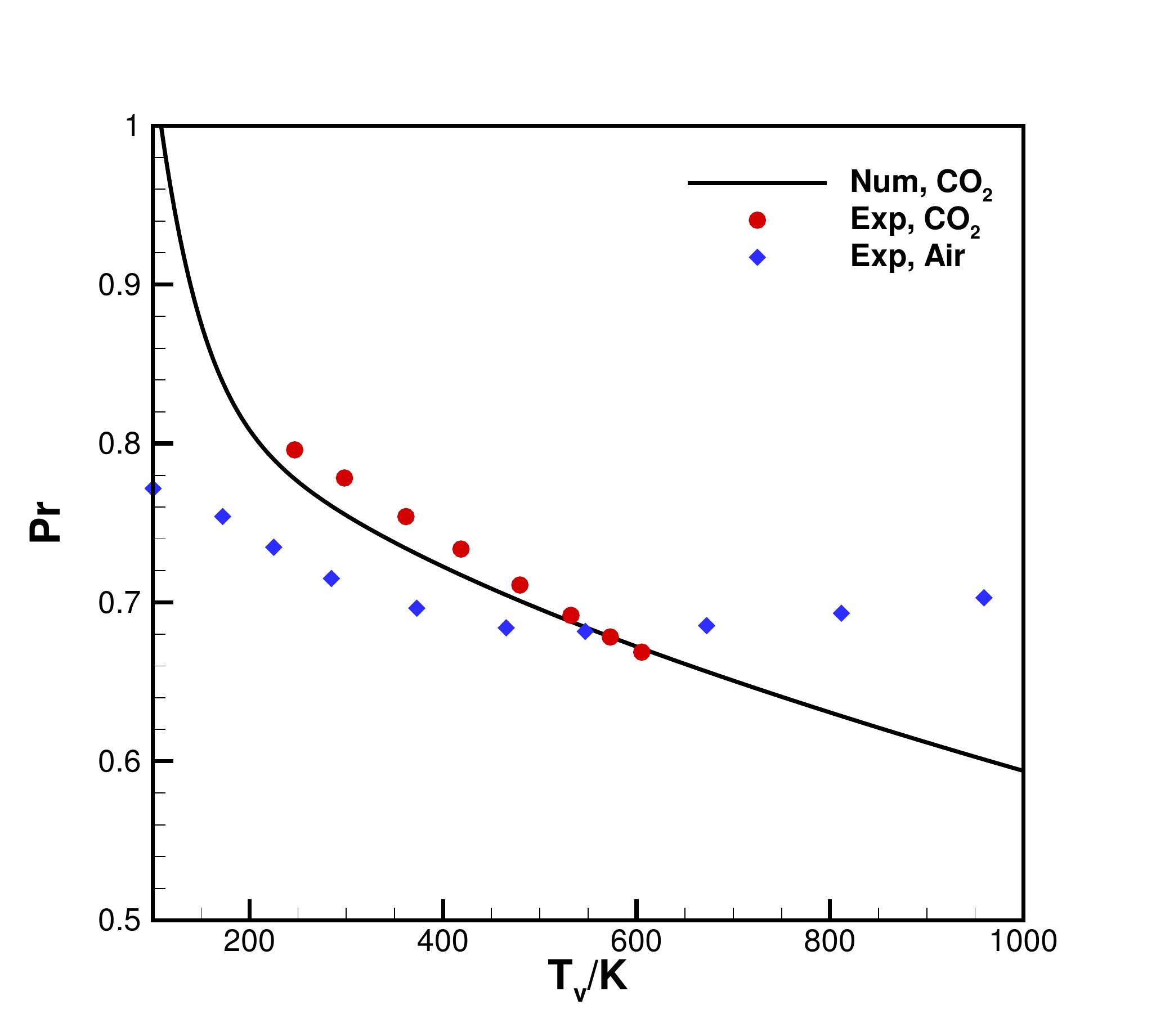}
	\vspace{-4mm}
	\caption{\label{gamma_pr} Comparisons on the specific heat ration $\gamma$ (left) and Prandtl number $Pr$ (right) for $CO_2$ and air. The numerical profile of $Pr$ is calculated with the assumption $T_t = T_v$. Experimental results are adapted from the Ref \cite{white2006viscous}. }
\end{figure}

For compressible wall-bounded turbulence, the Prandtl number $Pr$ plays a key role in determining the statistical turbulent quantities \cite{cao2021high}, especially for density and temperature fields.
For $CO_2$, the $Pr$ can be calculated as $Pr = \mu C_P/\kappa$, where shear viscosity $\mu$ and thermal conductivity $\kappa$ depends on the translational temperature $T_t$ as subsequent Eq.\eqref{shear_co2_2} and Eq.\eqref{thercon_co2_2}, and $C_P$ relies on the vibrational temperature as Eq.\eqref{gamma_co2}.
Figure \ref{gamma_pr} presents the comparisons on $\gamma$ and $Pr$ between $CO_2$ and air.
For air, the specific heat ratio $\gamma$ almost keeps as $1.4$ up to $800K$, and $Pr$ is fixed at approximate $0.7$ between $300K$ and $800K$.
In terms of $CO_2$, experimental measurements on $\gamma$ and $Pr$ show the strong temperature-dependent behavior \cite{white2006viscous}. 
We find the numerical profile of $\gamma$ in $CO_2$ with Eq.\eqref{gamma_co2} agrees well with the experimental one.
The numerical profile of $Pr$ is calculated with the assumption of $T_t = T_v$, and this assumption dose not hold for experimental measurement. 
Thus, it is reasonable to find the discrepancy in $Pr$ between the numerical profile and experimental result.
In following simulation, without the assumption of $T_t = T_v$ as shown in Figure \ref{gamma_pr}, the $Pr$ of $CO_2$ depends on the practical translational and vibrational temperatures.
Notice that the $\gamma$ and $Pr$ must be computed locally in each time step for each computational grid \cite{cai2008one}.

\subsection{Translational, rotational and vibrational relaxation time}
For $CO_2$ gas flows, non-equilibrium translational, rotational, and  vibrational internal modes are equipped with different relaxation time.
Aiming to the following construction of thermal non-equilibrium three-temperature BGK-type model, the translational, rotational and vibrational relaxation time requires to be determined firstly.
Translational relaxation time $\tau_t$ is related with shear viscosity $\mu$, and the power law \cite{white2006viscous} gives the approximation of shear viscosity as
\begin{align}\label{shear_co2_2}
	\mu(T_t) = \mu_0 (\frac{T_t}{T_0})^{n_t},
\end{align}
where $T_t$ is the translational temperature, $n_t = 0.79$ and $\mu_0 = 1.370 \times 10^{-5} kg/(m \cdot s)$ at $T_0 = 273 K$ for $CO_2$ is approximately valid between $209 K$ and $1700 K$.
While, $n_t = 0.666$ and $\mu_0 = 1.716 \times 10^{-5} kg/m \cdot s$ for air is approximately valid between $210 K$ and $1900 K$.
From Chapman-Enskog expansion \cite{chapman1970mathematical, xu2015direct}, the translational relaxation time $\tau_t$ of $CO_2$ can be obtained by $\tau_t = \mu/p$, where $p$ is the pressure.
Thus, the corresponding power law of translational relaxation time $\tau_t$ reads
\begin{equation}\label{tau_t_co2}
	\begin{aligned}
		\tau_t = \tau_{t0} (\frac{T_t}{T_0})^{n_t},
	\end{aligned}
\end{equation}
where $\tau_{t0} = 1.35 \times 10^{-10} s$ at $T_0 = 273 K$ with the atmosphere pressure $101.325 \times 10^3 Pa$.
The power law also gives the approximation of thermal conductivity $\kappa$ \cite{white2006viscous} as
\begin{equation}\label{thercon_co2_2}
	\kappa(T_t) = \kappa_0 (\frac{T_t}{T_0})^{n_{\kappa}},
\end{equation}
where $n_{\kappa} = 1.30$ and $\kappa_0 = 1.46 \times 10^{-2} W/(m \cdot K)$ at $T_0 = 273 K$ for $CO_2$ is approximately valid between $180 K$ and $700 K$.
While, $n_{\kappa} = 0.81$ and $\kappa_0 = 2.41 \times 10^{-2} W/(m \cdot K)$ for air is valid between $210 K$ and $2000 K$.
The thermal conductivity in Eq.\eqref{thercon_co2_2} gets involved with the calculation of $Pr$ as shown in Figure \ref{gamma_pr} and following numerical simulation in $CO_2$ turbulence.
For comparison, the Sutherland law for shear viscosity and thermal conductivity of $CO_2$ and air, as well as the curve-fit expression in the shear viscosity of $CO_2$ \cite{candler1990computation} over a much wider range of translational temperature has been provided in Appendix A.

The rotational relaxation time $\tau_r$ and vibrational relaxation time $\tau_v$ are related with the bulk viscosity $\eta_b$ \cite{cramer2012numerical}, which is given by
\begin{equation}\label{bulk_all}
	\eta_b = \eta_{b, c} + \eta_{b, int},
\end{equation}
where $\eta_{b, c}$ represents the effect of elastic collisions, and $\eta_{b, int}$ accounts for the inelastic collisional contribution of the internal d.o.f. \cite{emanuel1990bulk}. 
For dilute gas, $\eta_{b, c}$ can be neglected, resulting in
\begin{equation}\label{bulk_b}
\begin{aligned}
	\eta_b &\approx \eta_{b, int} 
	       = (\gamma - 1)^2 \sum_{i = 1}^{N} \frac{C_{V, l}}{R} p \tau_l,
\end{aligned}	
\end{equation}
where $\gamma$ is the specific heat ratio as Eq.\eqref{gamma_co2}, $R$ the gas constant, $C_{V, l}$ the specific heat at constant volume as Eq.\eqref{Cv_co2_t} - Eq.\eqref{Cv_co2_v}, and $\tau_l$ the relaxation time for the $l$th internal energy mode.
Assuming that the rotational and vibrational modes relax independently with a single rotational relaxation time $\tau_r$ and vibrational relaxation time $\tau_v$, the bulk viscosity as Eq.\eqref{bulk_b} can be rewritten as
\begin{align}
    \eta_b &= \eta_{b}^r + \eta_{b}^v, \label{bulk_b_rv}
\end{align}
with 
	\begin{align}
		\eta_{b}^r &= (\gamma - 1)^2 \frac{N_r}{2} p \tau_r, \label{bulk_b_r}\\ 
		\eta_{b}^v &= (\gamma - 1)^2 \frac{C_{V, v}}{R}  p \tau_v. \label{bulk_b_v}
	\end{align}
$\eta_{b}^r$ denotes the componential bulk viscosity arises from the rotational modes, and $\eta_{b}^v$ the componential bulk viscosity results from the vibrational modes.

The power law of $CO_2$ for rational relaxation time is calibrated by Eq.\eqref{bulk_b_r}, with the combination of experimental data \cite{wang2019bulk} and Parker equation \cite{parker1959rotational, kustova2021investigation}.
As reported in experimental study \cite{wang2019bulk}, the vibrational modes remain frozen when measuring the bulk viscosity of $CO_2$ with Rayleigh-Brillouin light scattering spectroscopy at $532nm$.
In such circumstance, the rotational modes contribute solely to the bulk viscosity, and Eq.\eqref{bulk_b_rv} reduces to $\eta_{b} = \eta_{b}^r$.
Thence, the seminal estimated experimental data of $\eta_{b}^r$ using Hammond-Wiggings hydrodynamic model \cite{wang2019bulk} is utilized to calibrate the $\tau_r$. Power law of rotational relaxation time $\tau_r$ is given by
\begin{equation}\label{tau_r_co2}
	\begin{aligned}
		\tau_r = \tau_{r0} (\frac{T_t}{T_0})^{n_r},
	\end{aligned}
\end{equation}
where $n_r = 1.59$ and $\tau_{r0} = 2.99 \times 10^{-10} s$ at $T_0 = 273 K$ is approximately valid between $250 K$ and $2000 K$ with the atmosphere pressure.
The details in calibrating the rotational relaxation time $\tau_r$ can be find in Appendix B.
With the well-known calibration of $p \tau_v$ \cite{cramer2012numerical}, the vibrational relaxation time can be  calibrated as 
\begin{equation}\label{tau_v_co2}
	\begin{aligned}
		\tau_v = \tau_{v0} (\frac{T_t}{T_0})^{n_v},
	\end{aligned}
\end{equation}
where $n_v = -1.353$ and $\tau_{v0} = 1.89 \times 10^{-6} s$ at $T_0 = 800 K$ is approximately valid between $300 K$ and $1700 K$ with the atmosphere pressure.

\vspace{-2mm}
\begin{figure}[!htp]
	\centering
	\includegraphics[width=0.485\textwidth]{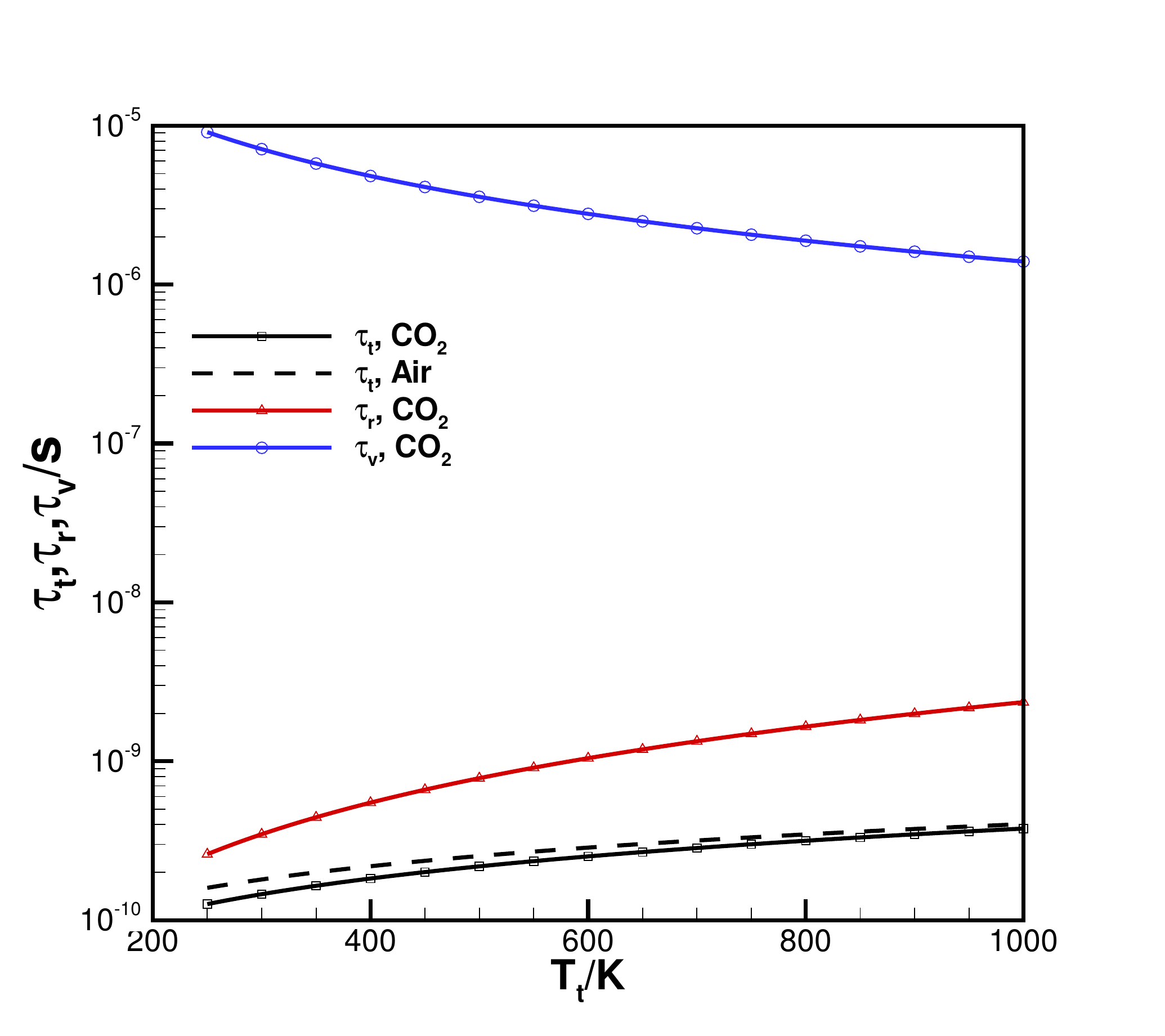}
	\vspace{-4mm}
	\caption{\label{fig_tau_trv_eta} Comparisons on translational relaxation time $\tau_t$, rotational relaxation time $\tau_r$ and vibrational relaxation time $\tau_v$ for $CO_2$ and air.}
\end{figure}
As power laws of Eq.\eqref{tau_t_co2}, Eq.\eqref{tau_r_co2} and Eq.\eqref{tau_v_co2} present, all relaxation times only depend on the translational temperature.
Figure \ref{fig_tau_trv_eta} presents the comparisons on translational, rotational and vibrational relaxation time for $CO_2$ and air.
In Figure \ref{fig_tau_trv_eta}, the translational relaxation time of air is slightly larger than that of $CO_2$.
We clearly observe that the vibrational relaxation time $\tau_v$ can be approximately thousands of times longer than that of $\tau_t$ and $\tau_r$.
Current quantitative calibration on $\tau_t$, $\tau_r$ and $\tau_v$ deviates from the previous theoretical calculations using kinetic theory \cite{kustova2019models}, while the qualitative behaviors are similar.
It should be noted that the bulk viscosity of $CO_2$ is mainly dominated by the $\eta_b^v$ arising from the vibrational modes \cite{wang2019bulk}.
Thus, present $\tau_v$ indeed gives the large bulk viscosity with Eq.\eqref{bulk_b_v}, which is consistent with the findings of inherent large bulk viscosity of $CO_2$ \cite{tisza1942supersonic, emanuel1990bulk}.

\section{Three-temperature kinetic model and numerical scheme for carbon dioxide}
In this section, an extended thermal non-equilibrium three-temperature BGK-type model for $CO_2$ is introduced with well-calibrated relaxation time.
The three-temperature kinetic equation is going to be solved by the proposed finite-volume non-equilibrium high-accuracy GKS.

\subsection{Three-temperature BGK-type model}
In terms of one-temperature equilibrium gas flow, the BGK model \cite{bhatnagar1954model} has been well proposed.
For thermal non-equilibrium three-temperature diatomic gas flow, the extended BGK-type model \cite{cai2008one, wang2017unified} reads
\begin{equation} 
	\begin{aligned}
	\frac{\partial f}{\partial t} + u_i \frac{\partial f}{\partial x_i} = \frac{f^{v} - f}{\tau_t} + \frac{f^{r} - f^v}{\tau_r} + \frac{g - f^{r}}{\tau_v} \equiv \frac{f^{v} - f}{\tau_t} + Q_v,
	\label{boltamann_bgk_neq}
\end{aligned}
\end{equation}
where $f(\boldsymbol{x}_{i+1/2, j_m, k_n}, t, \boldsymbol{u}, \xi_r, \xi_v)$ is the number density of molecules at position $(x_1, x_2, x_3)^T$ and time $t$, with particle velocity $\boldsymbol{u} = (u_1, u_2, u_3)^T$ and internal energy $(\xi_r, \xi_v)^T$.
The left hand side of Eq.\eqref{boltamann_bgk_neq} represents the free streaming of molecules in space, and the right side denotes the
collision term.
In Eq.\eqref{boltamann_bgk_neq}, two intermediate equilibrium states $f^{v}$ and $f^r$, and Maxwellian distribution $g$ are introduced with three temperatures $T_t$, $T_r$ and $T_v$ as
	\begin{align}	
	f^{v} &= \rho (\frac{\lambda_t}{\pi})^{\frac{N_t}{2}} e^{- \lambda_t (u_i - U_i)^2} (\frac{\lambda_r}{\pi})^{\frac{N_r}{2}} e^{- \lambda_r \xi_r^2} 
	(\frac{\lambda_v}{\pi})^{\frac{N_v}{2}} e^{- \lambda_v \xi_v^2}, \label{intermediate_f_v} \\
	f^{r} &= \rho (\frac{\lambda_t}{\pi})^{\frac{N_t}{2}} e^{- \lambda_t (u_i - U_i)^2} (\frac{\lambda_r}{\pi})^{\frac{N_r}{2}} e^{- \lambda_r \xi_r^2}, \label{intermediate_f_r} \\
	g &= \rho (\frac{\lambda_t}{\pi})^{\frac{N_t}{2}} e^{- \lambda_t (u_i - U_i)^2}, \label{intermediate_f_t}
\end{align}
where $\rho$ is the density, $\bm{U}$ denotes three-dimension velocities $(U_1, U_2, U_3)^T$, $\lambda_t = m_0/(2k_BT_t)$ is related to the translational temperature $T_t$, $\lambda_r = m_0/(2k_BT_r)$ and $\lambda_v = m_0/(2k_BT_v)$ account for the rotational temperature $T_r$ and vibrational temperature $T_v$, respectively. 
Above extended BGK-type kinetic model has the similarity with the two relaxation time BGK models for gases with internal d.o.f. \cite{morse1964kinetic, struchtrup1999bgk}.
For triatomic molecule $CO_2$, we reasonably adopt above three-temperature extend BGK-type model, with the well-calibrated relaxation time as Eq.\eqref{tau_t_co2}, Eq.\eqref{tau_r_co2} and Eq.\eqref{tau_v_co2}.
The right-hand-side collision operator as Eq.\eqref{boltamann_bgk_neq}  contains three terms, corresponding to the elastic collision ($f \to f^v$) and inelastic collisions ($f^v \to f^r$ and $f^r \to g$).
As shown in Figure \ref{fig_tau_trv_eta}, for multiple relaxation precess of $CO_2$, notice that the inelastic collision takes longer time than that of elastic collision.
The additional term $Q_v$ resulting from the inelastic collisions accounts for the internal energy exchange among the translational, rotational and vibrational internal energy.

The relation between density $\rho$, momentum $\rho \bm{U}$, total energy $\rho E$, rotational internal energy $\rho E_r$, and internal vibrational energy $\rho E_v$ are determined by taking moments of the intermediate equilibrium distribution function $f^v$ as
\begin{equation}
	\begin{aligned}
	\bm{Q} \equiv
	\int \bm{\psi_v} f^v \text{d} \Xi_v
	=
	\begin{pmatrix}
		\rho, \
		\rho \bm{U}, \
		\rho E, \
		\rho E_r, \
		\rho E_v
	\end{pmatrix}^T,
	\label{macro_vars}
\end{aligned}
\end{equation}
with the vector of extended collision invariants $ \bm{\psi_v} = (1, u_1, u_2, u_3, \frac{1}{2}(u_1^2 + u_2^2 + u_3^2 + \xi_r^2 + \xi_v^2), \frac{1}{2} \xi_r^2, \frac{1}{2} \xi_v^2)^T$ and $\text{d} \Xi_v = \text{d} u_1 \text{d} u_2 \text{d} u_3 \text{d} \xi_r \text{d} \xi_v$. 
In Eq.\eqref{macro_vars}, total energy is $E = \bm{U}^2/2 + E_t + E_r + E_v$.
Eq.\eqref{boltamann_bgk_neq} introduces the new rotational and vibrational temperatures $T_r$ and $T_v$, thus, the constraints of rotational and vibrational internal energy relaxation have to be imposed on the  extended kinetic model to self-consistently determine all unknowns.
Since only mass, momentum and total energy are conserved during molecule collisions, the compatibility condition for the collision term turns into
\begin{equation}\label{sourceterms_vector}
	\begin{aligned}
	\bm{S} \equiv \int (\frac{f^v - f}{\tau_t} +  Q_v) \bm{\psi_v} \text{d} \Xi_v = (0, 0, 0, 0, 0, S_r, S_v)^T.
\end{aligned}
\end{equation}
Source terms $S_r$ and $S_v$ are from the internal energy exchange among translational, rotational and vibrational modes during inelastic collision. 
These source terms cannot be derived from the BGK model itself.
These two source terms for the rotational internal energy and vibrational internal energy can be modeled through the Landau-Teller-Jeans-type relaxation model \cite{cai2008one,liu2021multiple}, which read
\begin{align}
	S_r &=  \frac{(\rho E_r)^{eq} - \rho E_r}{ \tau_r}, \label{sourceterms_Sr} \\
	S_v &=  \frac{(\rho E_v)^{eq} - \rho E_v}{ \tau_v}. \label{sourceterms_Sv}
\end{align}
For $CO_2$, the rotational and vibrational relaxation time $\tau_r$ and $\tau_v$ have been determined by Eq.\eqref{tau_r_co2} and Eq.\eqref{tau_v_co2}.
The left unknown equilibrium rotational internal energy $(\rho E_r)^{eq}$ and vibrational one $(\rho E_v)^{eq}$ are determined by the assumption $T_v = T_r = T_t = T^{eq}$ \cite{cai2008one, liu2021multiple}, i.e.,
	\begin{align}
	T_{eq} &\equiv \frac{N_t T_t + N_r T_r + N_v T_v}{N_t + N_r + N_v} \label{Temperature_eq}, \\
	(\rho E_r)^{eq} &= \frac{N_r}{2} \rho R T_{eq}, \label{Energy_eq_r}\\ 
	(\rho E_v)^{eq} &= \frac{N_v}{2} \rho R T_{eq}, \label{Energy_eq_v}
\end{align}
with the translational d.o.f. as $N_t$, rotational d.o.f. as $N_r$ and vibrational d.o.f. as $N_v$. 
Up to this point, the thermal non-equilibrium three-temperature BGK-type model for $CO_2$ is completed with the well-determined relaxation time $\tau_t$, $\tau_r$, and $\tau_v$, as well as the modeling source terms $S_r$ and $S_v$.

Using the intermediate equilibrium state $f^v$, with the frozen of rotational and vibrational internal energy exchange, the $1$st-order Chapman-Enskog expansion \cite{chapman1970mathematical} gives
\begin{equation}
	\begin{aligned}
	f = f^v - \tau_t (\frac{\partial{f^v}}{\partial t} + u_i \frac{\partial f^v}{\partial x_i}),
	\label{ce_expansion_neq}
\end{aligned}
\end{equation}
from which the corresponding non-equilibrium three-temperature macroscopic governing equations in three-dimensions can be derived \cite{liu2021multiple}. 
The technical details in the standard derivation of three-temperature macroscopic equations are similar as that of deriving two-temperature macroscopic equations \cite{cao2021highphd}, which are omitted in this paper.
The key finding is that two additional equations get involved with above source terms Eq.\eqref{sourceterms_Sr} and Eq.\eqref{sourceterms_Sv}, which govern the evolutionary dynamics of rotational internal energy $\rho E_r$ and vibrational internal energy $\rho E_v$.
Thus current thermal non-equilibrium three-temperature macroscopic system goes beyond the one-temperature supersonic turbulent channel flow of thermally perfect gas \cite{chen2019effects}.
The non-equilibrium three-temperature macroscopic equations give the fixed Prandtl number $Pr = 1$. 
In the numerical simulation of supersonic $CO_2$ turbulent channel flow, the heat flux through the cell interface will be corrected to obtain the targeted $Pr$ to any realistic value \cite{white2006viscous}.
The three-temperature BGK-type equation as Eq.\eqref{boltamann_bgk_neq} is going to be solved by following non-equilibrium high-accuracy GKS, as the numerical fluxes at cell interfaces are evaluated based on the time-dependent gas distribution solution.

\subsection{Non-equilibrium high-accuracy gas-kinetic scheme}\label{subsection_HGKSn}
For finite-volume non-equilibrium GKS, the key procedure is updating the macroscopic flow variables inside each control volume through the numerical fluxes. 
In this section, the spatial and temporal high-accuracy non-equilibrium GKS is proposed within the two-stage fourth-order framework \cite{pan2016efficient}.

Taking moments of the extended three-temperature BGK-type model as Eq.\eqref{boltamann_bgk_neq} and integrating with respect to control volume, the finite volume scheme can be expressed as
\begin{equation}\label{semi}
	\begin{aligned}
		\frac{\text{d}(\boldsymbol{Q}_{ijk})}{\text{d}t}=-\frac{1}{|\Omega_{ijk}|}\sum_{s=1}^6\mathbb{F}_{s}(t) + \boldsymbol{S}_{ijk},
	\end{aligned}
\end{equation}
where $\boldsymbol{Q}_{ijk}$ is the cell averaged macroscopic variables as Eq.\eqref{macro_vars}, $\boldsymbol{S}_{ijk}$ is the cell averaged source term as Eq.\eqref{sourceterms_vector}.
The control volume $\Omega_{ijk}=[(x_1)_i-\Delta x_1/2, (x_1)_i+\Delta x_1/2] \boldsymbol{\cdot} [(x_2)_j - \Delta x_2/2, (x_2)_j+\Delta x_2/2] \boldsymbol{\cdot} [(x_3)_k-\Delta x_3/2, (x_3)_k+\Delta x_3/2]$,
$|\Omega_{ijk}|$ is the volume of $\Omega_{ijk}$ and $\mathbb{F}_{s}(t)$ is the time-dependent numerical flux across the cell interface $\Sigma_{s}$. 
The numerical flux $\mathbb{F}_{s}(t)|_{x_1}$ in $x_1$- direction is given as example
\begin{equation}\label{flux_x}
	\begin{aligned}
		\mathbb{F}_{s}(t)|_{x_1}
		&=\iint_{\Sigma_{s}|_{x_1}}
		\boldsymbol{F} (\boldsymbol{Q}) \boldsymbol{\cdot} \boldsymbol{n} \text{d}\sigma \\
		&= \sum_{m, n = 1}^2\omega_{mn}
		\int \boldsymbol{\psi_v} u_1
		f(\boldsymbol{x}_{i+1/2, j_m, k_n}, t, \boldsymbol{u}, \xi_r, \xi_v)\text{d}\Xi_v \Delta x_2 \Delta x_3,
	\end{aligned}
\end{equation}
where $\boldsymbol{n}$ is the outer normal direction. 
The Gaussian quadrature is used over the cell interface for Eq.\eqref{flux_x}, where $\omega_{mn}$ is the quadrature weight,
$\boldsymbol{x}_{i+1/2, j_m, k_n}=[(x_1)_{i+1/2}, (x_2)_{j_m}, (x_3)_{k_n}]^T$, and $[(x_2)_{j_m}, (x_3)_{k_n}]$ is the quadrature point of cell
interface $[(x_2)_j - \Delta x_2/2, (x_2)_j+\Delta x_2/2] \boldsymbol{\cdot} [(x_3)_k-\Delta x_3/2, (x_3)_k+\Delta x_3/2]$. 
When constructing numerical fluxes for Eq.\eqref{boltamann_bgk_neq}, the secondary relaxation term $Q_v$ is splitly taken into account as the source term $\boldsymbol{S}_{ijk}$ in Eq.\eqref{semi}.

Without considering $Q_v$, the gas distribution function $f(\boldsymbol{x}_{i+1/2, j_m, k_n}, t, \boldsymbol{u}, \xi_r, \xi_v)$ in the local coordinate can be obtained by the integral solution of Eq.\eqref{boltamann_bgk_neq} as 
\begin{equation} \label{formal_solution}
	\begin{aligned}
		f(\boldsymbol{x}_{i+1/2, j_m, k_n}, t, \boldsymbol{u}, \xi_r, \xi_v) 
		= \frac{1}{\tau_t}\int_0^t
		f^v(\boldsymbol{x}', t', \boldsymbol{u}, \xi_r, \xi_v)e^{-(t-t')/ \tau_t}\text{d}t'
		+e^{-t/\tau_t}f_0(-\boldsymbol{u}t, \xi_r, \xi_v),
	\end{aligned}	
\end{equation}
where $\boldsymbol{x}'=\boldsymbol{x}_{i+1/2, j_m, k_n}-\boldsymbol{u}(t-t')$
is the trajectory of molecule on grids, $f_0$ the initial gas
distribution function, and $f^v$ the corresponding intermediate equilibrium state in the form of Eq.\eqref{intermediate_f_v}. 
Along the line of GKS \cite{xu2001gas}, for the multi-dimensional kinetic solver, $f^v$ and $f_0$ can be constructed as
\begin{equation}
	\begin{aligned}
		f^v = f^v_0(1 + \overline{a}_1 x_1 + \overline{a}_2 x_2 + \overline{a}_3 x_3 + \overline{A} t),
	\end{aligned}
\end{equation}
and
\begin{equation}
	\begin{aligned}
		f_0 =
		\begin{cases}
			f^v_l [1 +  (a_1^l x_1 + a_2^l x_2 + a_3^l x_3) - \tau_t (a_1^l u_1 + a_2^l u_2 + a_3^l u_3 + A_l)], &x \leq 0, \\
			f^v_r [1 +  (a_1^r x_1 + a_2^r x_2 + a_3^r x_3) - \tau_t (a_1^r u_1 + a_2^r u_2 + a_3^r  u_3 + A_r)], &x > 0,
		\end{cases}
	\end{aligned}
\end{equation}
where $f^v_l$ and $f^v_r$ are the initial gas distribution functions on both sides of a cell interface $\Sigma_{s}$.
$f^v_0$ is the initial intermediate equilibrium state located at the cell interface, which can be determined through the compatibility condition
\begin{align}
	\int \boldsymbol{\psi_v} f^v_0 \text{d}\Xi= \int_{u_1>0} \boldsymbol{\psi_v} f^v_l \text{d}\Xi_v + \int_{u_1<0} \boldsymbol{\psi_v} f^v_r \text{d}\Xi_v.
\end{align}
Substituting $f^v$ and $f_0$ into Eq.\eqref{formal_solution}, the time-dependent gas distribution function at the Gaussian point is evaluated as
\begin{equation}
	\begin{aligned}
		\label{formalsolution_neq}
		f(\boldsymbol{x}_{i+1/2, j_m, k_n}, t, \boldsymbol{u}, \xi_r, \xi_v) 
		&= (1 - e^{-t/\tau_t}) f^v_0 + ((t + \tau_t) e^{-t\tau_t} - \tau_t) (\overline{a}_1 u_1 + \overline{a}_2 u_2 + \overline{a}_3 u_3) f^v_0   \\
		&+ (t - \tau_t + \tau_t e^{-t\tau}) \overline{A} f^v_0  \\
		&+ e^{-t/\tau_t} f^v_l [1 - (\tau_t + t) (a_1^l u_1 + a_2^l u_2 + a_3^l u_3) - \tau_t A_l] H(u_1) \\
		&+ e^{-t/\tau_t} f^v_r [1 - (\tau_t + t) (a_1^r u_1 + a_2^r u_2 + a_3^r u_3) - \tau_t A_r] (1 - H(u_1)).
	\end{aligned}
\end{equation}
With the relation of macroscopic variables and intermediate equilibrium distribution function $f^v$, the spatial mesoscopic coefficients $\overline{a}_1$, $a_1^l$, $\cdots$, $a_3^l$, $a_3^r$ and temporal mesoscopic coefficients $\overline{A}$, $A_l$, $A_r$ in Eq.\eqref{formalsolution_neq} can be determined and details are presented in Appendix C. 
It is noticed that Eq.\eqref{formalsolution_neq} provides a gas evolution process from kinetic scale to hydrodynamic scale on grids, where both inviscid and viscous fluxes are recovered from a time-dependent and multi-dimensional gas distribution function at a cell interface.
This flux function couples the inviscid and all dissipative terms \cite{xu2001gas, cao2018physical}, and has advantages in comparison with traditional hydrodynamic solver in which the Riemann solver \cite{toro2013riemann} and central difference are used for the inviscid and viscous terms splitly.
For Prandtl number fix, similar to Ref \cite{xu2001gas}, the total energy flux $\boldsymbol{F}{(\rho E)}$ in Eq.\eqref{flux_x} is modified as 
$\boldsymbol{F}^{new} {(\rho E)}= \boldsymbol{F}{(\rho E)} + (1/Pr - 1) q$, 
where the time-dependent heat flux through the cell interface can be evaluated precisely by 
$q = \int (u - U) \{[(u_i - U_i)^2 + \xi_r^2 + \xi_v^2]/2 \} f d\Xi_v$.
As presented in Section \ref{subsection_rvm}, $Pr$ depends on both the translational and vibrational temperatures, which is computed locally in each time step for each computational grid.

With the time-dependent kinetic flux as Eq.\eqref{formalsolution_neq}, the second-order accuracy can be achieved by one step integration. 
To achieve high-order accuracy in space and time, the fifth-order WENO-Z spatial reconstruction \cite{castro2011high} and two-stage fourth-order time discretization \cite{li2016two, pan2016efficient} are implemented. 
For source term in Eq.\eqref{semi}, the one-step forward Euler method is applied in two-stage updating process to guarantee the robustness.
Notice that the $T_v$ and $N_v$ as Eq.\eqref{Etrv_co2_v} and Eq.\eqref{Nv_co2} should be calculated in the center of control volume for source term, as well as at the cell interface when calculating non-equilibrium fluxes in Appendix C.
Newton–Raphson method is utilized to compute the $T_v$ at $n+1$ step.
With the initial guess $T_v$ at $n$ step, several iterations are enough to obtain the convergent solution, i.e., with convergence error $\epsilon_{T_v} \equiv ||T_v^{*} - T_v^{**}||/T_v^n \le 10^{-8}$ where $T_v^{*}$ and $T_v^{**}$ are the successive iterated value during the iteration process.
$||\phi||$ denotes the absolute value of $\phi$.
The present non-equilibrium high-accuracy GKS has been constructed with the second-order kinetic flux as Eq.\eqref{formalsolution_neq}, fifth-order WENO-Z reconstruction, two-stage fourth-order time discretization.
Current non-equilibrium high-accuracy GKS is well implemented in the in-house parallel computational platform for turbulence simulation \cite{cao2021three, cao2022high}, and the DNS in supersonic $CO_2$ turbulent channel flow is presented subsequently.

\section{Numerical simulation and discussion}
Numerical simulations in supersonic turbulent channel flows with bulk Mach number $Ma = 3.0$ and bulk Reynolds number $Re = 4880$ are implemented in this section.
The benchmark as equilibrium one-temperature supersonic turbulent channel flow is validated with HGKS \cite{cao2021high, cao2022high} firstly. 
The supersonic thermal non-equilibrium three-temperature $CO_2$ turbulent channel flow is simulated with detailed physical analysis.

\subsection{One-temperature supersonic turbulent channel flow}\label{subsection_onetemp_validation}
The computational studies of supersonic turbulent channel flow
\cite{coleman1995numerical, chen2019effects, yu2019genuine, zhang2020contribution} have been extensively carried out to study the compressible turbulent boundary layer. 
One-temperature supersonic turbulent channel flow with bulk Mach number $Ma = 3.0$ and bulk Reynolds number $Re = 4880$ \cite{coleman1995numerical, zhang2020contribution} is firstly used to validate the high-accuracy of in-house HGKS solver with non-uniform grids \cite{cao2021high}. 
In the computation, the physical domain is
$(x, y, z)\in[0,4\pi H]\times[-H, H]\times[0,4\pi H/3]$ and the
computational domain takes $(\xi, \eta, \zeta)\in[0,4\pi H]\times[0,3\pi H]\times[0,4\pi H/3]$. 
The coordinate transformation is given as previous studies \cite{cao2021high, cao2022high}.
The periodic boundary conditions are used in streamwise $X$-direction and spanwise $Z$-directions, and the non-slip and isothermal boundary conditions are utilized in wall-normal $Y$-direction.
In what follows, note that $X$-, $Y$- and $Z$- directions are equivalent as $x_1$-,$x_2$- and $x_3$- directions in Section \ref{subsection_HGKSn}.
$(U_1, U_2, U_3)^T$ is re-expressed in $(U, V, W)^T$ for convenient comparisons with refereed studies.

The turbulent channel flow is initiated with uniform density $\rho = 1$, 
and the initial streamwise velocity $U(y)$ profile is given by the perturbed Poiseuille flow profile $U(y) = 1.5(1-y^{2}) + \text{white noise}$, where the white noise is added with $10\%$ amplitude of local streamwise velocity. 
The spanwise and wall-normal velocity is initiated with white noise. 
The initial uniform pressure is computed through the corresponding intial bulk Mach number and Reynolds number. 
The targeted non-dimensional parameters bulk Mach number $Ma$ and bulk Reynolds number $Re$ are defined as $Ma = U_b/c_w$, $Re = \rho_b U_bH / \mu_w$.
The bulk velocity $U_b$ and bulk density $\rho_b$ are given by
$U_b =  \int_{-H}^{H} U(y) \text{d}y$ and $\rho_b =  \int_{-H}^{H} \rho(y) \text{d}y$.
$H = 1$ is the half height of the channel, $\mu_w$ the wall
molecule viscosity.
$c_w = \sqrt{\gamma R T_w}$ is the wall sound speed, $T_w$ the wall temperature and $R$ the gas constant. 
$T_w$ is set to $1$ in current simulation.
In current on-temperature validation case, the shear viscosity $\mu$ adopts the exact same power law $\mu(T) \propto T^{0.7}$ \cite{coleman1995numerical}, where $T$ is the equilibrium temperature.
The plus unit $Y^+$ and plus velocity $U^+$ are defined as
$Y^+ = \rho u_\tau y/\mu$, $U^+ = U/u_{\tau}$ with the friction velocity $u_\tau = \sqrt{\tau_w/\rho_w}$, the wall shear stress $\tau_w = \mu_w \partial U /\partial y\big|_{w}$, and the wall density $\rho_w$.
The friction Mach number $Ma_\tau$ and the friction Reynolds
number $Re_{\tau}$ are given by $Ma_\tau = u_{\tau}/{c_w}$ and $Re_{\tau} = H/ \delta_v$ with $\delta_v =  \mu_w/(\rho_w u_{\tau})$.
The heat flux $q_{w}$ and the non-dimensional heat flux $B_q$ of the wall are defined as $q_{w} = - \kappa \partial T / \partial y \big|_{w}$,  $B_q = q_w / (\rho_w C_P u_{\tau} T_w)$.
In the one-temperature validation, the fixed Prandtl number $Pr = 0.70$ is used as refereed simulation \cite{coleman1995numerical}, which is close to the $Pr$ of air as shown in Figure \ref{gamma_pr}.
These statistical quantities are used to quantitatively validate the performance of HGKS and non-equilibrium high-accuracy GKS subsequently.

\begin{table}[!h]
	\centering
	\begin{tabular}{c|c|c|c|c|c}
		\hline \hline
		Case       &Physical domain  &$N_x \times N_y \times N_z$   &$\Delta Y^{+}_{min}$/$Y^{+}_{N10}$   &$\Delta X^{+}$   &$\Delta Z^{+}$    \\
		\hline
		Ref$_1$    &$4 \pi H \times 2H \times 4 \pi H/3$  &$144 \times 90 \times 60$    &0.20/17   &39   &24   \\
		\hline
		Ref$_2$    &$4 \pi H\times 2H \times 4 \pi H/3$  &$400 \times 210 \times 320$    &0.65/-      &14.32       &5.96    \\
		\hline
		$G_1$      &$4 \pi H\times 2H \times 4 \pi H/3$  &$128 \times 128 \times 128$    &0.52/12.94   &43.36    &14.45  \\
		\hline \hline
	\end{tabular}
	\caption{\label{cchannel_parameters} Supersonic one-temperature turbulent channel flow: numerical parameters of current validation case $G_1$ and the reference simulations \cite{coleman1995numerical, zhang2020contribution}. "-" means that the data can not be find in the refereed paper.}
\end{table}
\begin{figure}[!h]
	\centering
	\includegraphics[width=0.485\textwidth]{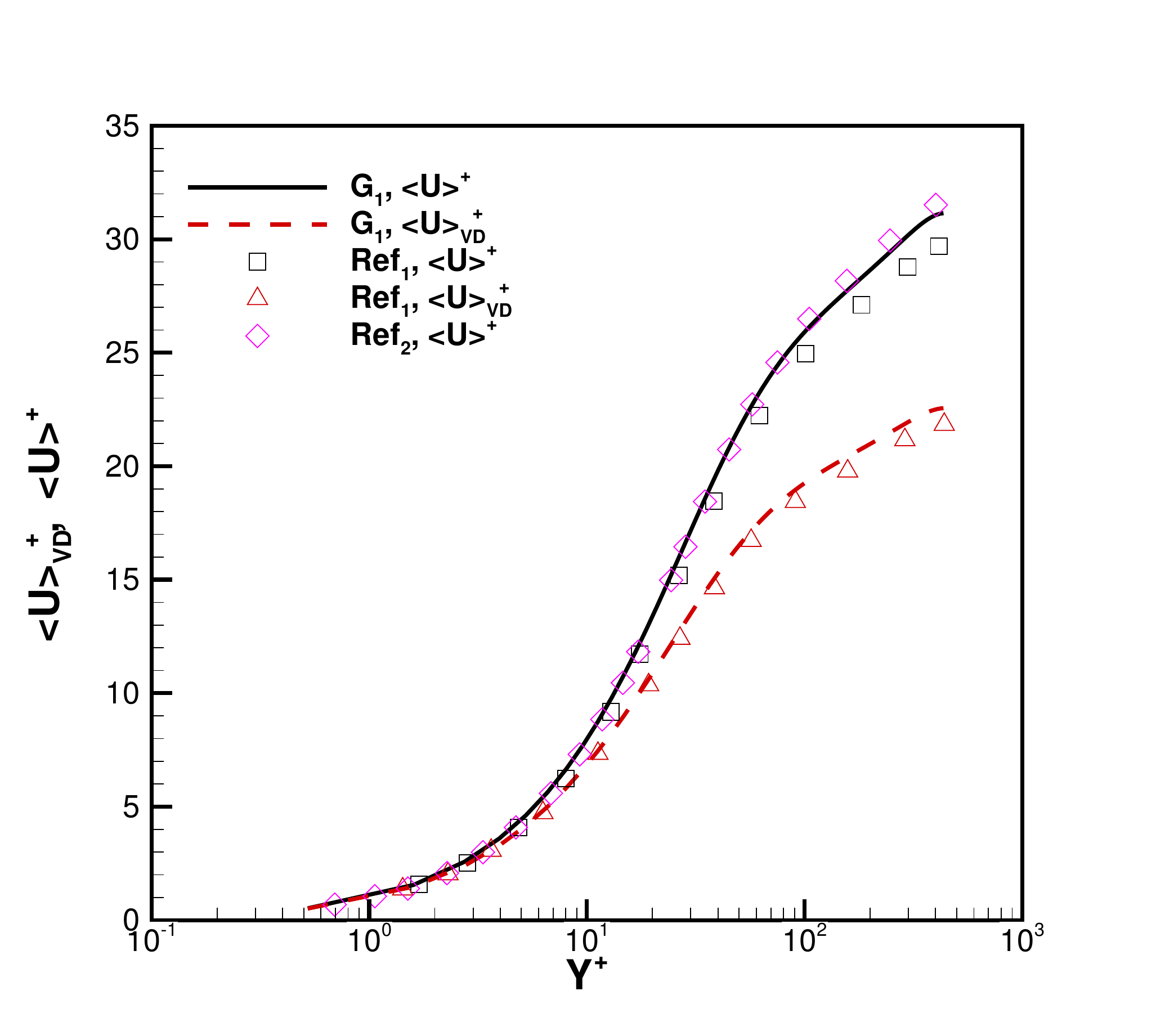}
	\includegraphics[width=0.485\textwidth]{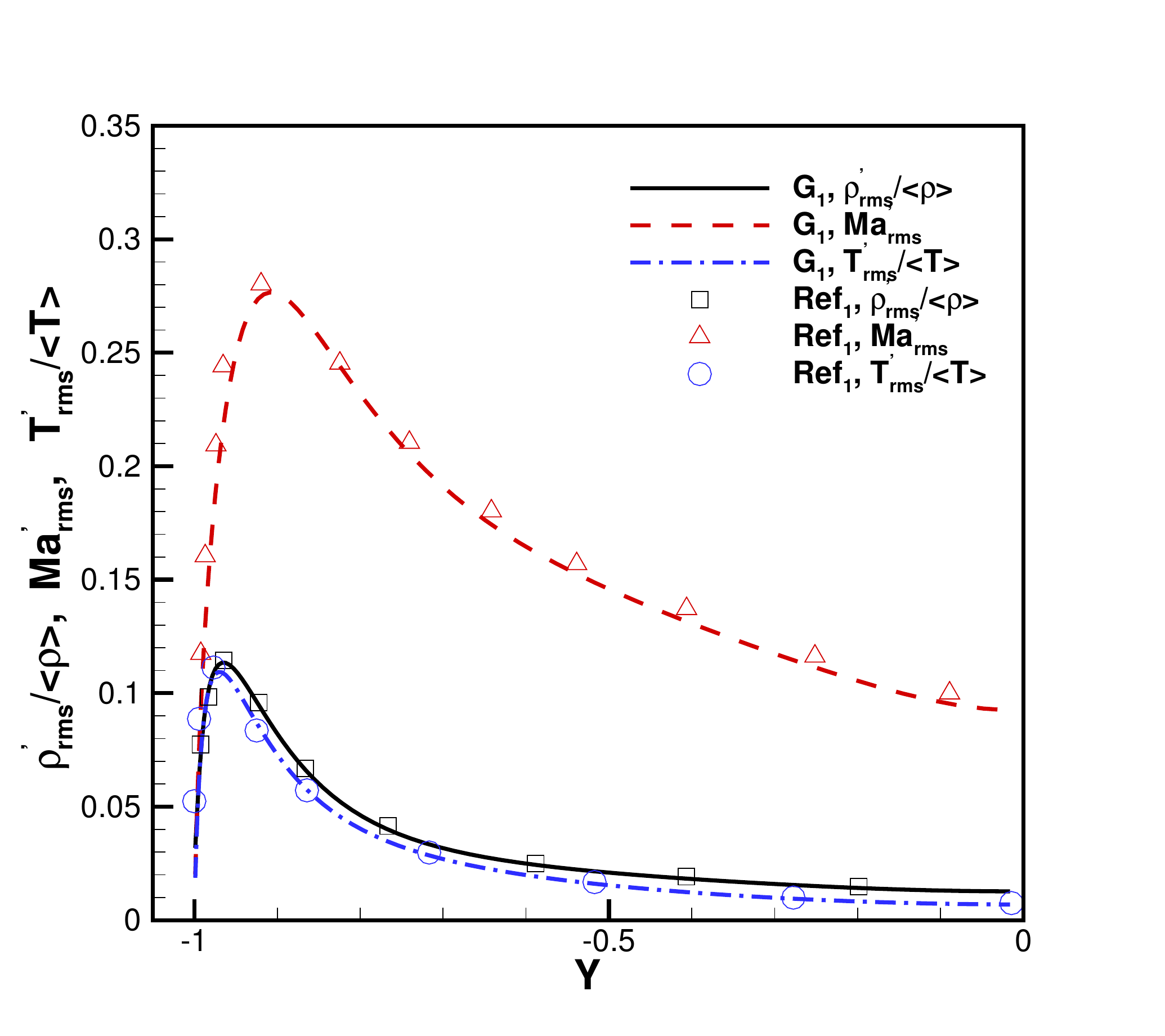}	
	\includegraphics[width=0.485\textwidth]{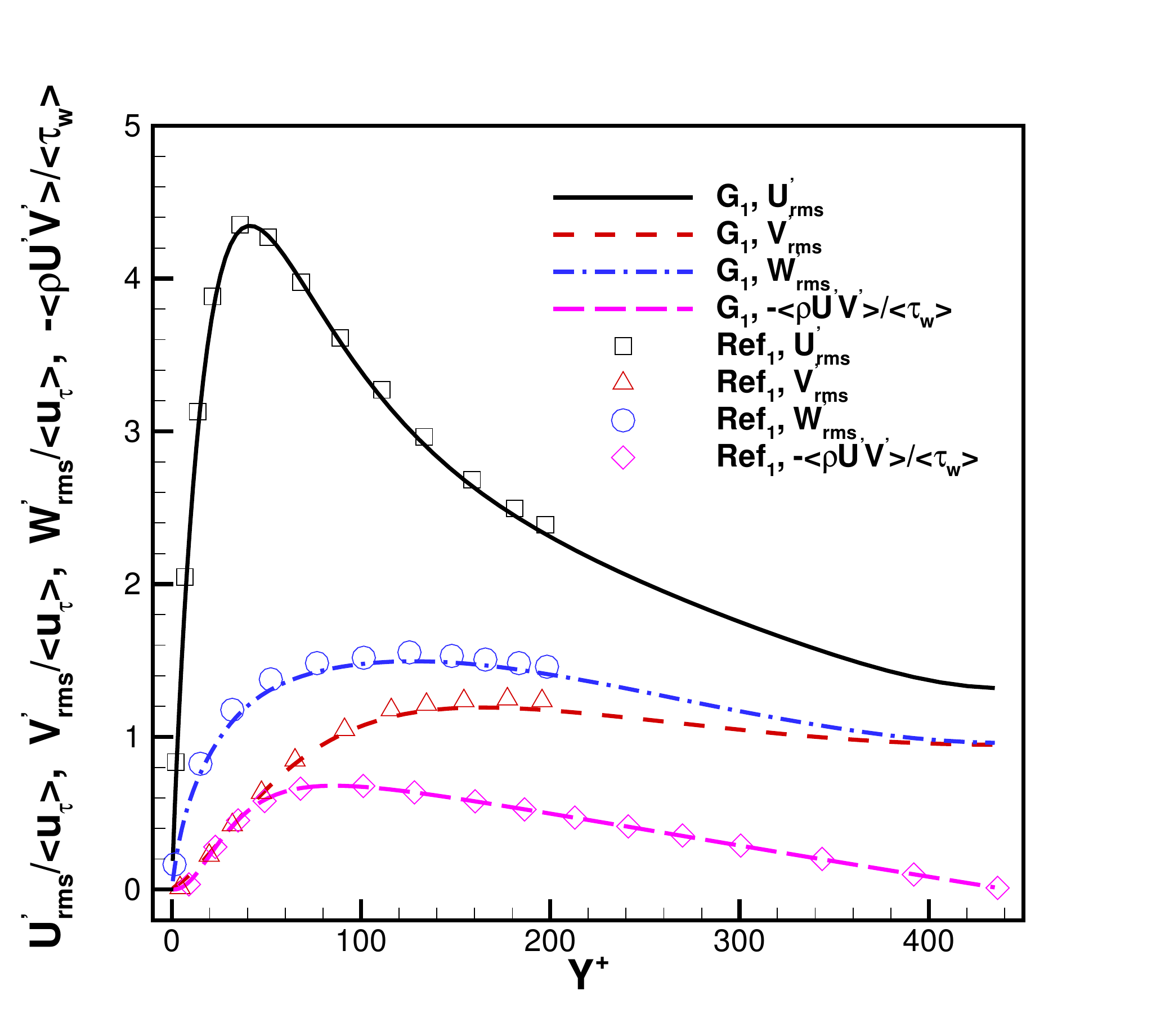}
	\vspace{-4mm}
	\caption{\label{cchannel_rhotma_rms} Supersonic one-temperature turbulent channel flow: the profiles of normalized r.m.s. of density $\rho_{rms}^{'}/ \langle \rho \rangle$, r.m.s. of Mach number $Ma_{rms}^{'}$ and normalized r.m.s. of temperature $T_{rms}^{'} / \langle T \rangle$ (upper), and the profiles of normalized turbulence intensities $U_{rms}^{'}/\langle u_{\tau} \rangle$, $V_{rms}^{'}/\langle u_{\tau} \rangle$, $W_{rms}^{'}/\langle u_{\tau} \rangle$ and Reynolds stress $-\langle \rho U^{'} V^{'} \rangle/\langle \tau_{w} \rangle$ (lower).}
\end{figure}
In this one-temperature supersonic turbulent channel flow simulation, the details of numerical parameters are given in Table \ref{cchannel_parameters}. 
The numerical cases of DNS in refereed paper\cite{coleman1995numerical}
and \cite{zhang2020contribution} are denoted as Ref$_1$ and Ref$_2$, and case $G_1$ is implemented by one-temperature HGKS \cite{cao2022high}.
The spectral method and high-order difference scheme is used in Ref$_1$ and Ref$_2$, respectively.
$\Delta Y^{+}_{min}$ is the first grid center space
off the wall in the wall-normal direction, and $Y^{+}_{N10}$ is the
plus unit for the first ten points (grid center) off the wall. 
$\Delta X^{+}$ and $\Delta Z^{+}$ are the equivalent plus unit for uniform streamwise and spanwise grids, respectively.
As shown in Table \ref{cchannel_parameters}, the grid resolution of case $G_1$ meet the requirement for DNS \cite{coleman1995numerical}.
The constant moment flux is used to determine the external force \cite{cao2022high} in transition and the fully developed turbulence periods. 
The supersonic turbulent channel flow takes longer time to transit than that of near incompressible turbulent channel flow with $Ma = 0.1$.
During the fully developed turbulence period, $680$ characteristic periodic time $H/U_b$ is used to obtain the statistically stationary turbulence. 
In what follows, the ensemble average of $\phi$ over time and the $X$- and $Z$-directions is represented by $\langle \phi \rangle$.
The fluctuation of $\phi$ is denoted by $\phi^{'} = \phi - \langle \phi \rangle$, and the root-mean square (r.m.s.) of $\phi$ is defined as $\phi_{rms}^{'} = \sqrt{ \langle (\phi - \langle \phi \rangle)^2 \rangle}$, where $\phi$ represents the density, temperature and velocity, etc. 
To further quantify the performance of HGKS in one-temperature supersonic turbulent channel flow, the profiles of normalized r.m.s. of density $\rho_{rms}^{'} / \langle \rho \rangle$, r.m.s. of Mach number $Ma_{rms}^{'}$, and normalized r.m.s. temperature $T_{rms}^{'} / \langle T \rangle$, and the profiles of normalized turbulence intensities (r.m.s. of velocities as $U_{rms}^{'}/\langle u_{\tau} \rangle$, $V_{rms}^{'}/\langle u_{\tau} \rangle$, $W_{rms}^{'}/\langle u_{\tau} \rangle$) and Reynolds stress $-\langle \rho U^{'} V^{'} \rangle/\langle \tau_{w} \rangle$ are presented in Figure \ref{cchannel_rhotma_rms}. 
In order to account for the mean property of variations caused by
compressibility, the Van Driest (VD) transformation \cite{huang1994van} is proposed for the mean velocity, i.e., density-weighted velocity
${\left\langle U \right\rangle}_{VD}^+ = \int_{0}^{{\left\langle U \right\rangle}^+} ( \left\langle \rho \right\rangle / \left\langle \rho_w \right\rangle)^{1/2} \text{d} {\left\langle U \right\rangle}^+$.
Compared with the refereed DNS solution \cite{coleman1995numerical}, the well-agreed performance of case $G_1$ confirms the high-accuracy flow-fields has been obtained by current HGKS for one-temperature supersonic turbulent channel flow.
The small deviations between the case $G_1$ and Ref$_1$ may result from the numerical solutions in different governing equations.
For VD transformation of streamwise velocity, the solution from case $G_1$ matches well with the DNS in very fine grids \cite{zhang2020contribution} with fixed Prandtl number $Pr = 0.72$.
Subsequently, non-equilibrium high-accuracy GKS for three-temperature model proposed in Section \ref{subsection_HGKSn} is used to simulate supersonic thermal non-equilibrium three-temperature $CO_2$ turbulent channel flow.

\subsection{Three-temperature supersonic $CO_2$ turbulent channel flow}
For the first time, the DNS in supersonic thermal non-equilibrium three-temperature $CO_2$ turbulent channel flow is implemented.
The numerical setup is same as one-temperature validation case $G_1$ in Section \ref{subsection_onetemp_validation}.
The bulk Mach number and bulk Reynolds number take $Ma = 3.0$ and $Re = 4880$.
The three-temperature initial flow fields restarts from the one-temperature fully developed turbulence, with initializing three temperatures as $T_t = T_r = T_v$.
With absolute wall temperature $T_w = 1$, the equivalent wall temperature $T_{we} = 300K$ is adopted to compute the practical physical temperatures when determining $\tau_r$ and $\tau_v$. 
The isothermal boundary condition is utilized for translational temperature $T_t$.
To the author's knowledge, there is no report on wall boundary condition for $CO_2$.
In current study, assuming rotational and vibrational modes of $CO2$ do not exchange internal energy with the wall, thus, adiabatic boundary condition is used for rotational and vibrational temperatures in wall-normal $Y$-direction.
The realistic wall boundary conditions for rotational and vibrational internal energy of $CO_2$ deserve to be explored by seminal experimental measurements and theoretical studies.

\begin{figure}[!h]
	\centering
	\includegraphics[width=0.485\textwidth]{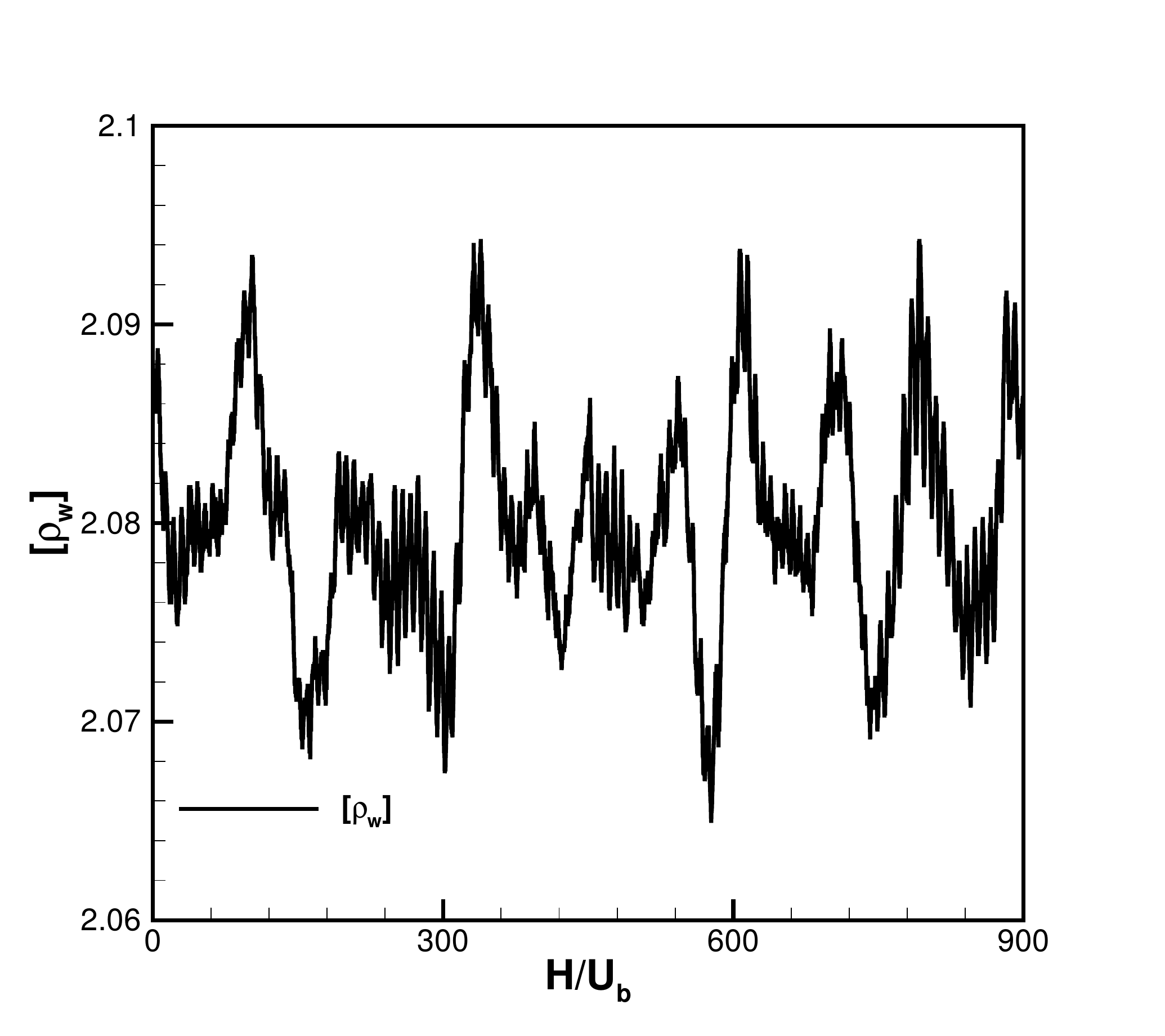}
	\includegraphics[width=0.485\textwidth]{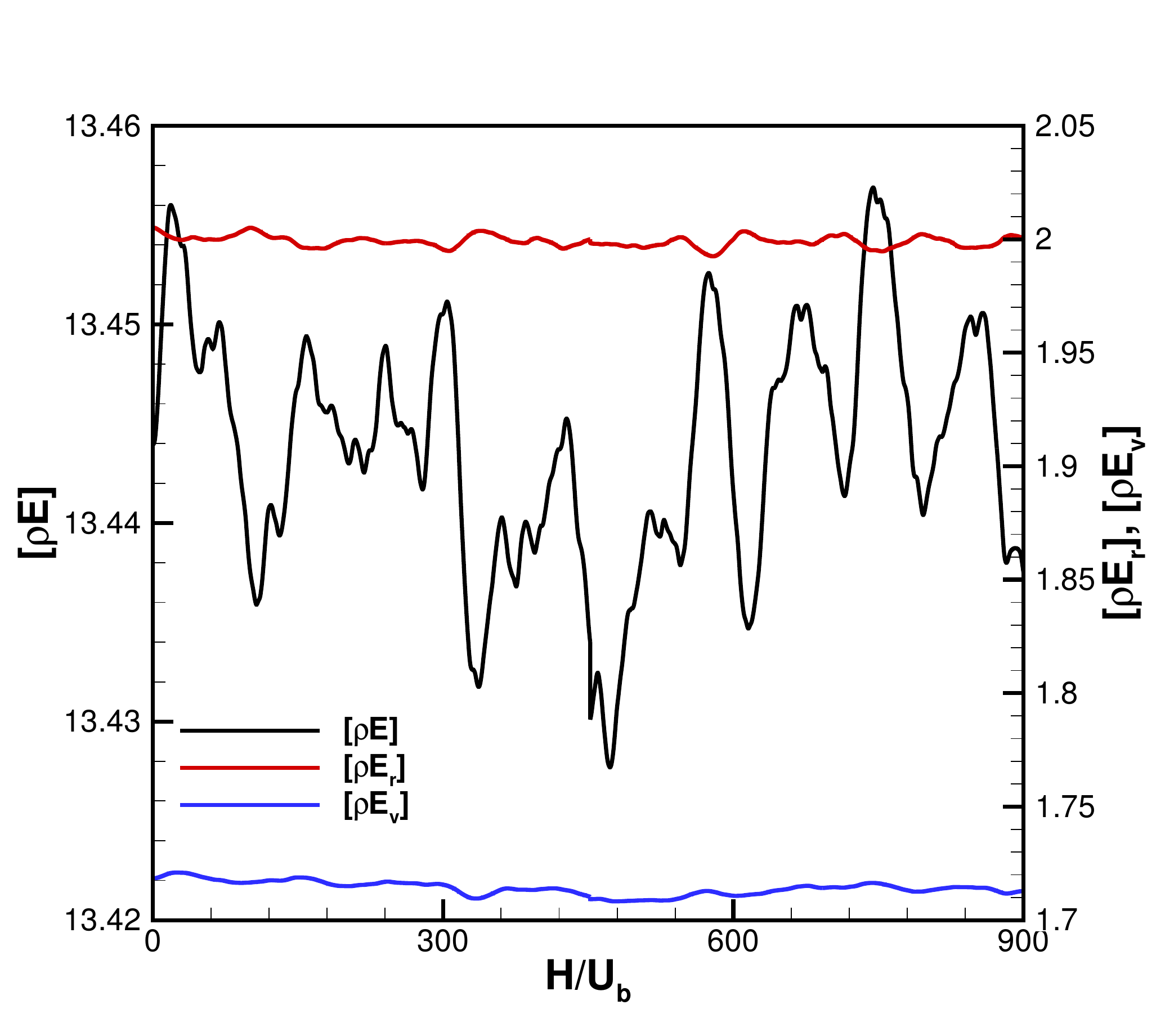}
	\vspace{-4mm}
	\caption{\label{co2_cchannel_rhowal_rhoE} Supersonic thermal non-equilibrium three-temperature $CO_2$  turbulent channel flow: the time history of mean wall density $[\rho_w]$ (left), and the time history of mean total energy $[\rho E]$, mean rotational internal energy $[\rho E_r]$ and mean vibrational internal energy $[\rho E_v]$ (right).}
\end{figure}
\begin{figure}[!h]
	\centering
	\includegraphics[width=0.495\textwidth]{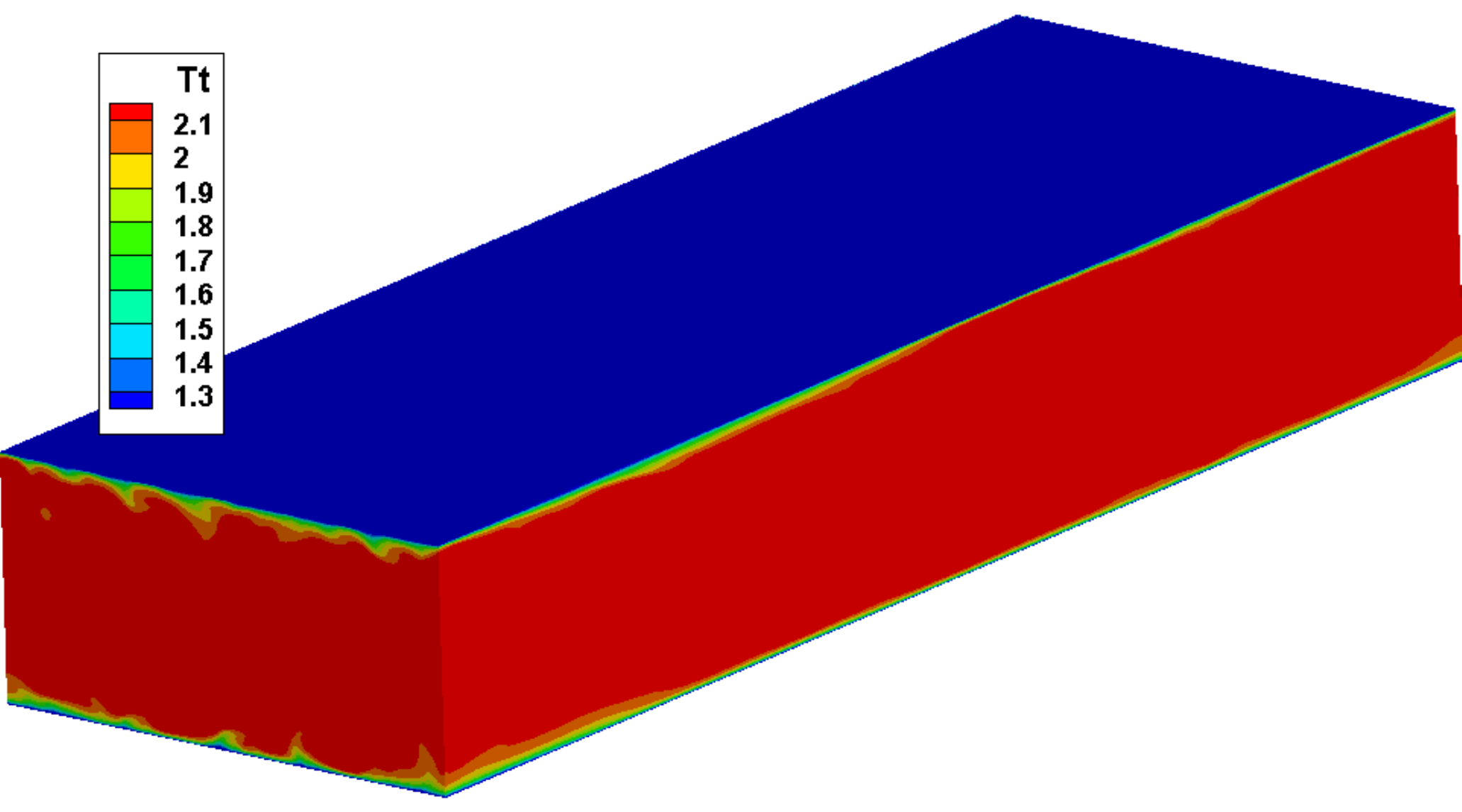}
	\includegraphics[width=0.495\textwidth]{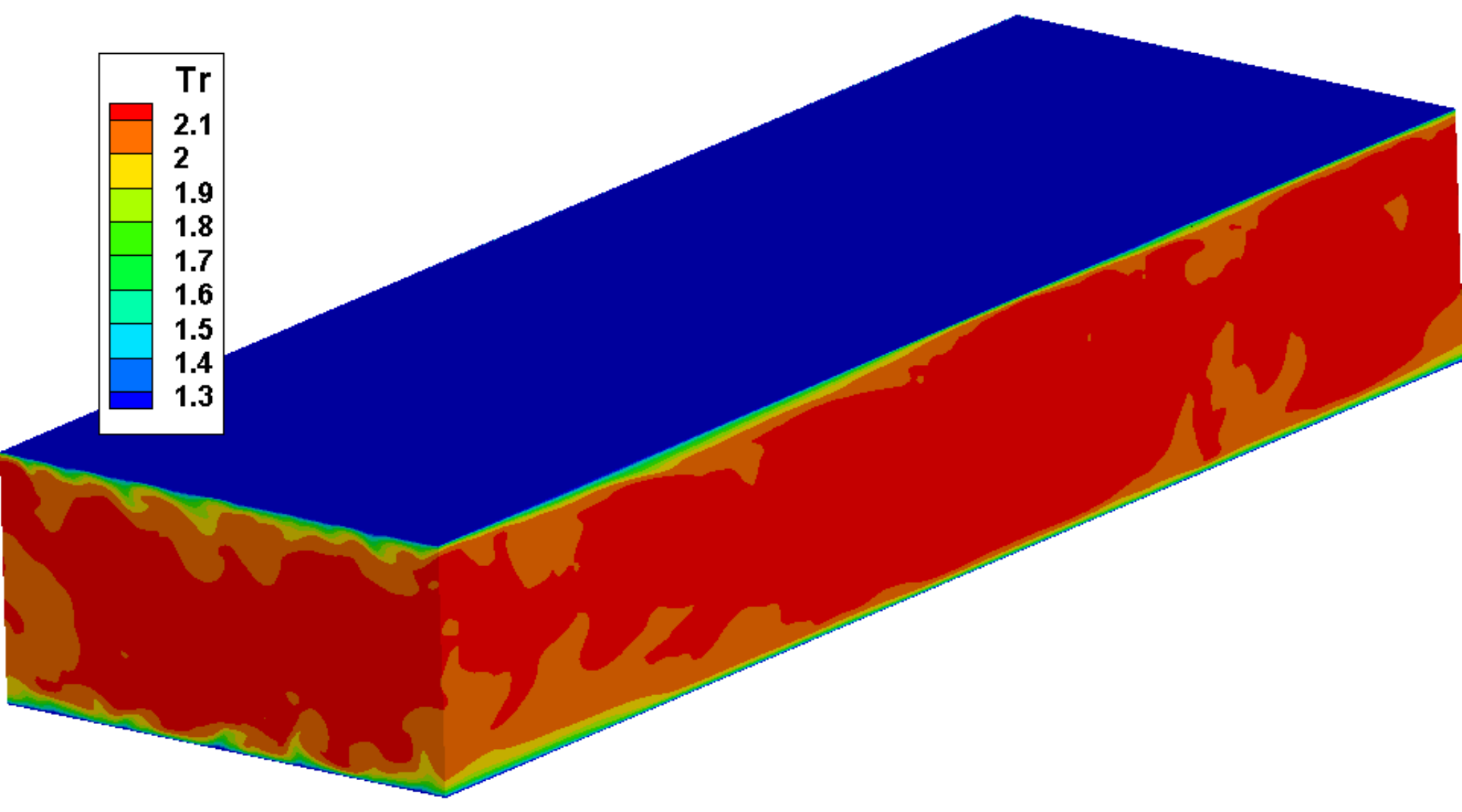}
	\includegraphics[width=0.495\textwidth]{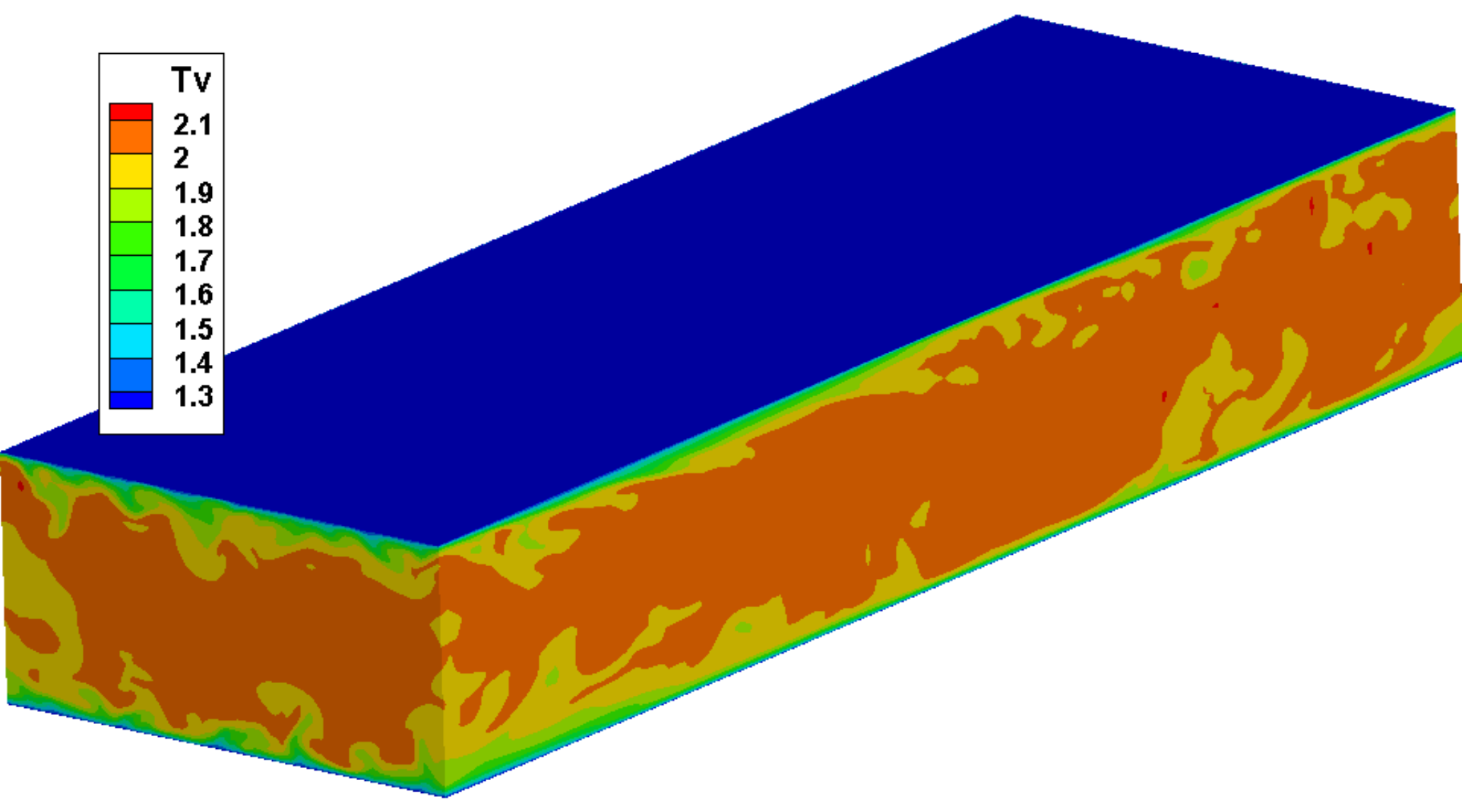}
	\caption{\label{co2_contours_3D3T} Supersonic thermal non-equilibrium three-temperature $CO_2$ turbulent channel flow: contours of translational, rotational, and vibrational temperatures.}
\end{figure}
To balance the wall shear stress, the constant moment flux is used to determine the external force of supersonic thermal non-equilibrium three-temperature $CO_2$ turbulent channel flow.
In Figure \ref{co2_cchannel_rhowal_rhoE}, note that the mean average of $\phi$ over whole computational domain (or the two wall planes) is represented by $[\phi]$.
After the long running time, the statistical mean variables as $[\rho_w]$, $[\rho E]$, $[\rho E_r]$ and $[\rho E_v]$ in Figure \ref{co2_cchannel_rhowal_rhoE} oscillate in a narrow range, indicating that the three-temperature $CO_2$ turbulent channel flow reaches the fully developed state.
Specifically, the mean total energy $[\rho E]$ of one-temperature case $G_1$ is approximate $94\%$ of current three-temperature $CO_2$. 
The larger mean total energy $[\rho E]$ of three-temperature $CO_2$ is reasonable, since the $CO_2$ turbulence is equipped with the additional excited vibrational internal energy.
Figure \ref{co2_contours_3D3T} shows three-dimension contours of translational, rotational, and vibrational temperatures of supersonic $CO_2$ turbulent flows in the fully developed turbulence period.
We clearly observe that the contour of vibrational temperature distinguishes from the translational and rotational temperature contours, confirming the thermal non-equilibrium performance.

\begin{table}[!h]
	\centering
	\begin{tabular}{c|c|c|c|c|c|c|c}
		\hline \hline
		Case          &${\left\langle \rho_w \right\rangle}$ &${\left\langle u_{\tau} \right\rangle}/U_b$   &${\left\langle Ma_{\tau} \right\rangle}$  &${\left\langle Re_{\tau} \right\rangle}$         &${\left\langle B_q \right\rangle}$   &${\left\langle \rho_c \right\rangle}$  &${\left\langle T_c^{eq} \right\rangle}$\\
		\hline
		Ref$_1$   &2.388    &0.0387     &0.116      &451         &0.137    &0.952    &2.490\\
		\hline 
		G$_1$     &2.407    &0.0376     &0.113      &442         &0.137    &0.948    &2.521\\
		\hline
		C$_1$     &2.080    &0.0400     &0.120      &406         &0.127    &0.966    &2.084\\		
		\hline \hline
	\end{tabular}
	\caption{\label{co2_cchannel_re_tau} Supersonic thermal non-equilibrium three-temperature $CO_2$ turbulent channel flow: ensemble quantities at the wall and the center plane of channel.}
\end{table}
\begin{figure}[!htp]
	\centering
	\includegraphics[width=0.485\textwidth]{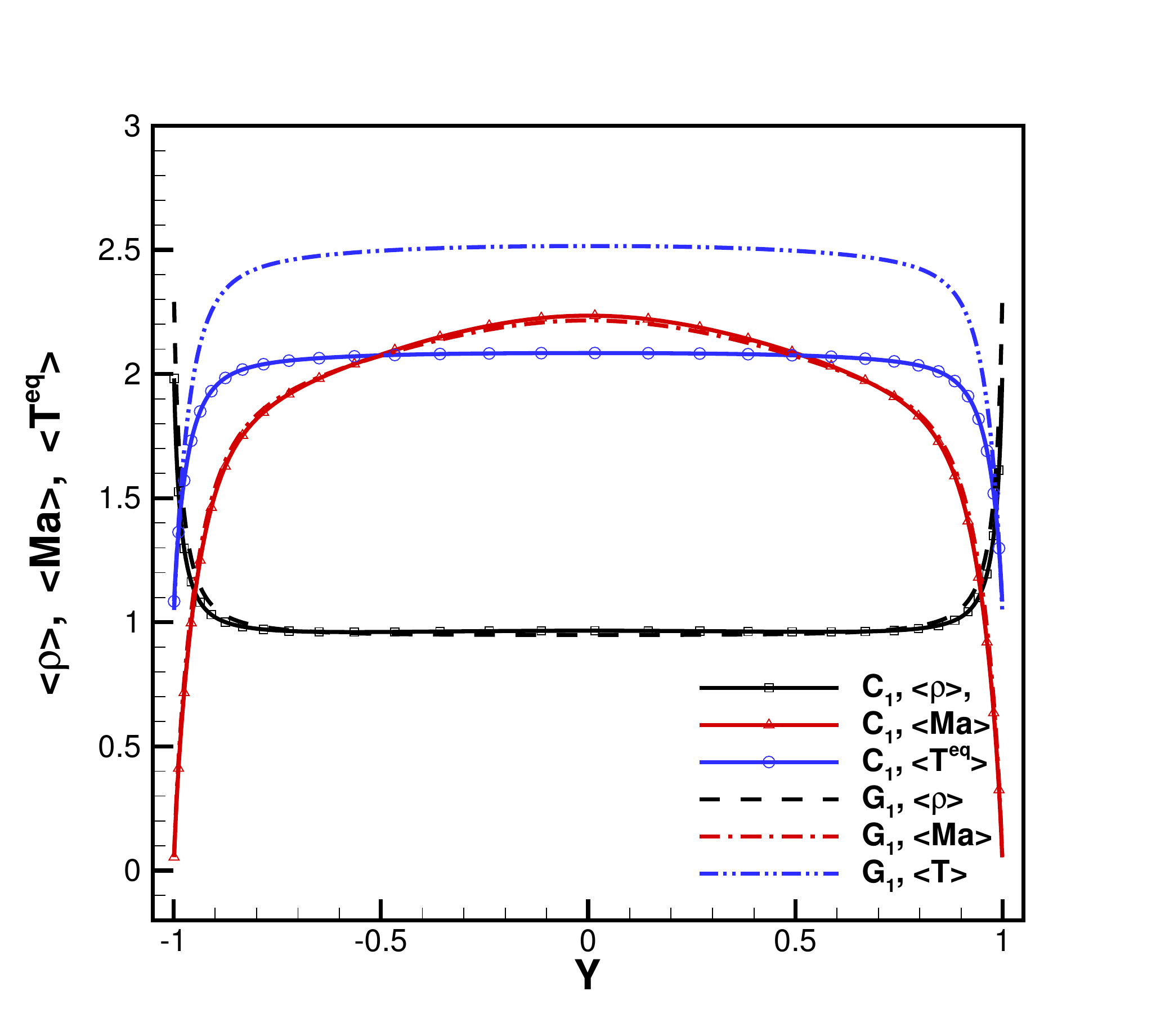}
	\includegraphics[width=0.485\textwidth]{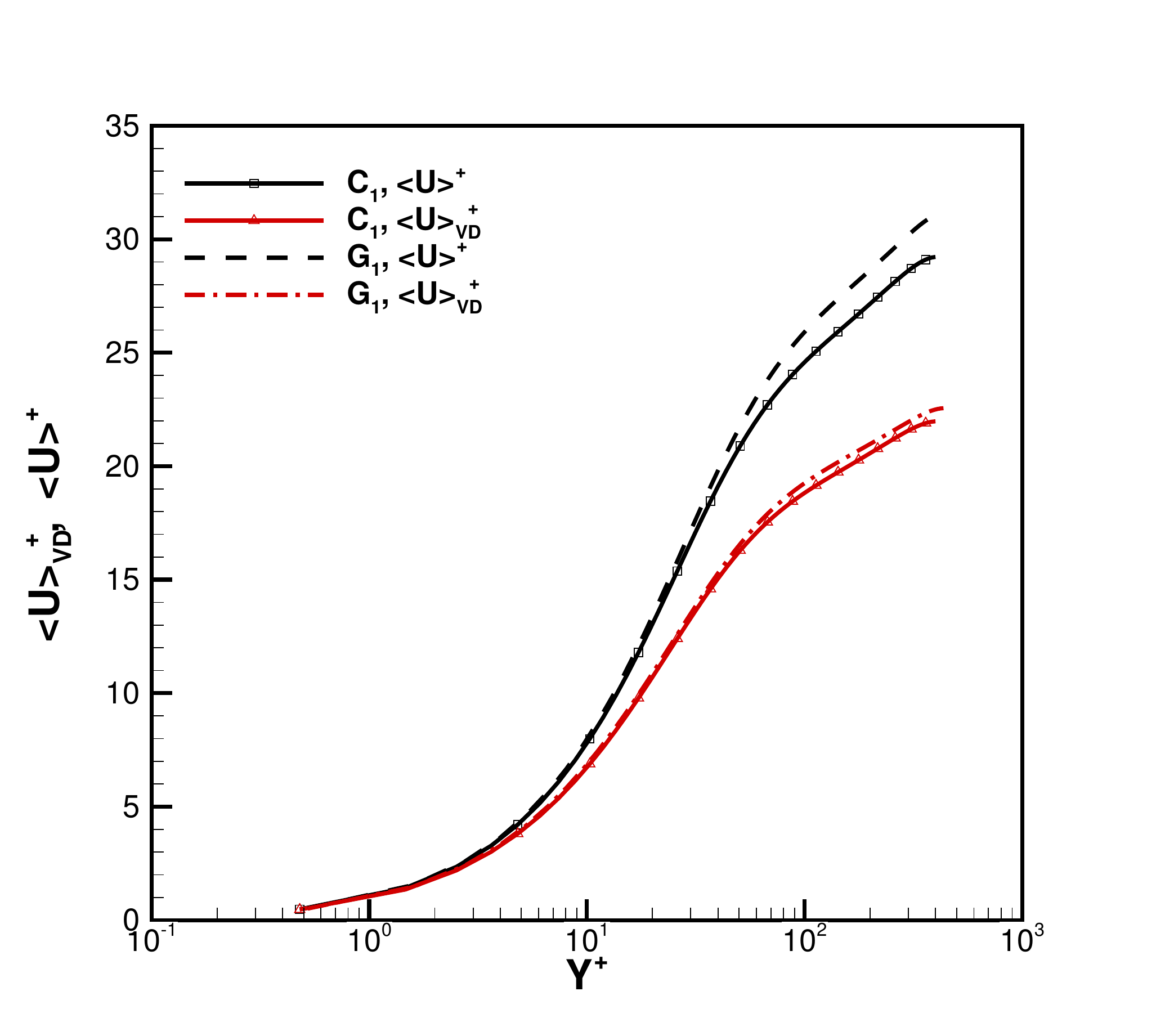}
	\vspace{-4mm}
	\caption{\label{co2_cchannel_rhotma_rms} Supersonic thermal non-equilibrium three-temperature $CO_2$ turbulent channel flow: the profiles of ensemble density $\langle \rho \rangle$, ensemble Mach number $\langle Ma \rangle$, and ensemble equilibrium temperature $\langle T^{eq} \rangle$ (left), and the profiles of ensemble streamwise velocity profiles ${\left\langle U \right\rangle}^+$ and VD transformation of streamwise velocity ${\left\langle U \right\rangle}_{VD}^+$ (right).}
\end{figure}
\begin{figure}[!h]
	\centering
	\includegraphics[width=0.485\textwidth]{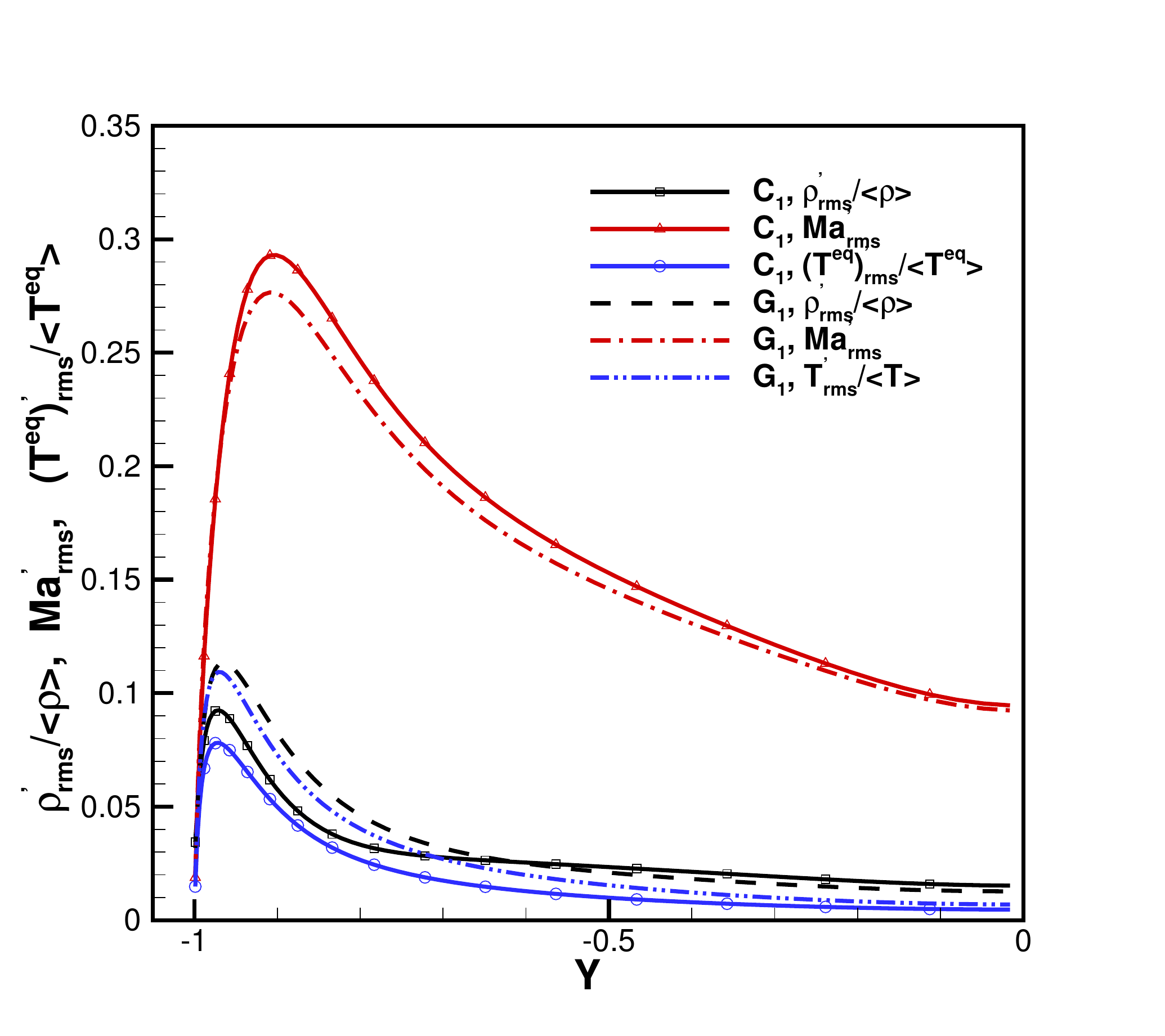}
	\includegraphics[width=0.485\textwidth]{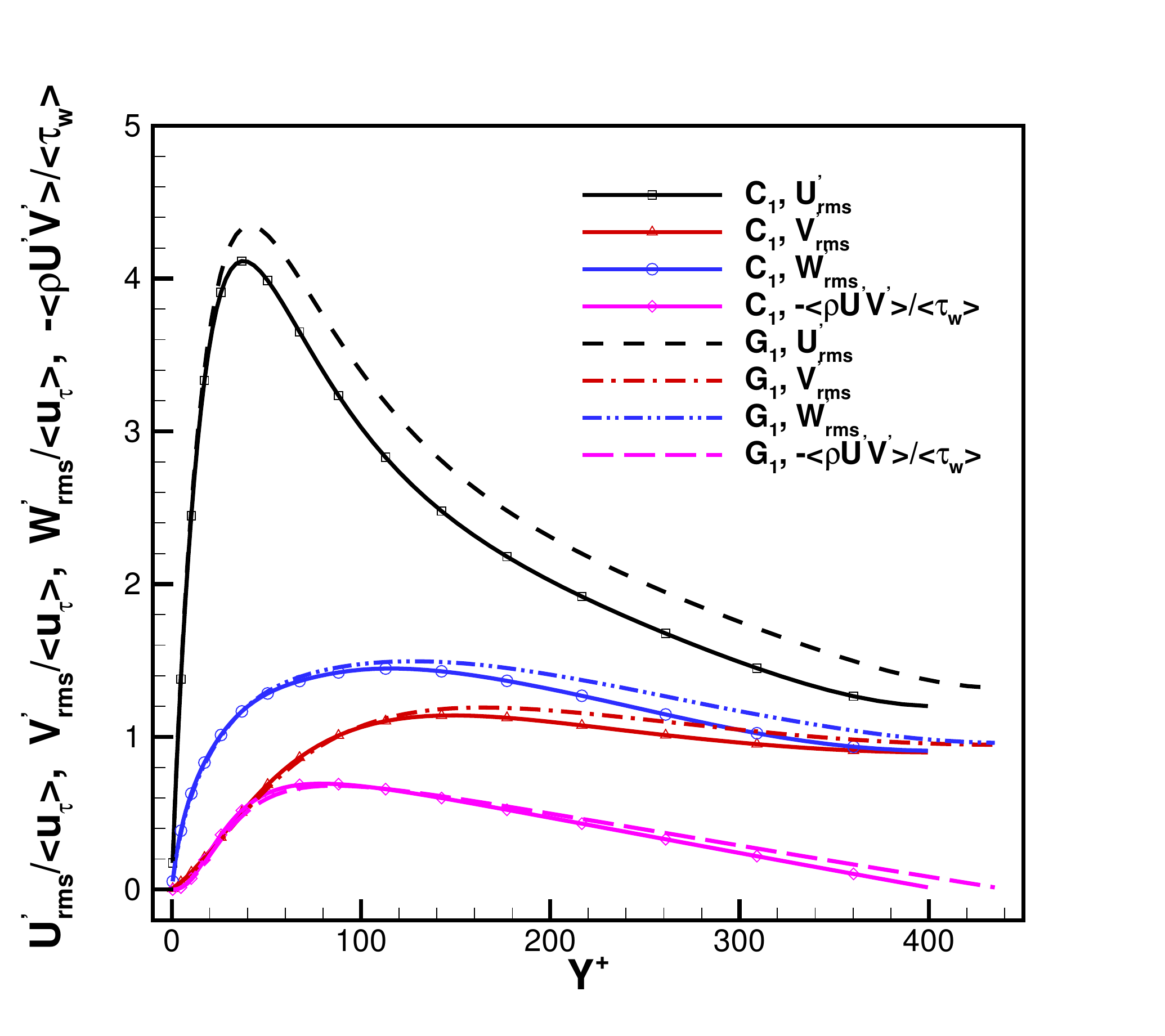}
	\vspace{-4mm}
	\caption{\label{co2_cchannel_uplusvdrms} Supersonic thermal non-equilibrium three-temperature $CO_2$ turbulent channel flow: the profiles of normalized r.m.s. of density $\rho_{rms}^{'}/ \langle \rho \rangle$, r.m.s. of Mach number $Ma_{rms}^{'}$ and normalized r.m.s. of equilibrium temperature ${(T^{eq})}^{'}_{rms}/ \langle T^{eq} \rangle$ (left), and the profiles of normalized turbulence intensities $U_{rms}^{'}/\langle u_{\tau} \rangle$, $V_{rms}^{'}/\langle u_{\tau} \rangle$, $W_{rms}^{'}/\langle u_{\tau} \rangle$ and normalized Reynolds stress $-\langle \rho U^{'} V^{'} \rangle/\langle \tau_{w} \rangle$  (right).}
\end{figure}
To quantitatively analyze the three-temperature performance of supersonic $CO_2$ turbulence, $900$ characteristic periodic time $H/U_b$ is used to obtain the statistically stationary three-temperature turbulence.
The key ensemble quantities at the wall and the center plane of channel are presented in Table \ref{co2_cchannel_re_tau}.
Running case $C_1$ denotes the supersonic thermal non-equilibrium three-temperature $CO_2$ case.
Again, ensemble quantities from case $G_1$ agree well with these of refereed solution in Ref$_1$ \cite{coleman1995numerical}.
Compared with the solutions of one-temperature case $G_1$, 
the ensemble friction velocity $\langle u_{\tau} \rangle$, the ensemble friction Mach number ${\left\langle Ma_{\tau} \right\rangle}$ and the ensemble center density ${\left\langle \rho_c \right\rangle}$ (density at the center plane of channel, namely at $Y=0$ plane) of case $C_1$ are slightly larger.
The non-dimensional heat flux $B_q$ of the wall of case $C_1$ is slightly smaller than that of case $G_1$.
While the ensemble wall density ${\left\langle \rho_w \right\rangle}$, the ensemble friction Reynolds number $Re_{\tau}$, and the ensemble central equilibrium temperature ${\left\langle T_c^{eq} \right\rangle}$ of case $C_1$ reduce dramatically.
Frictional force and heat transfer of the wall is of special interest in the long-term applicability of compressible $CO_2$ turbulence.
The ensemble frictional force  $\langle \tau_w \rangle$ of case $G_1$ and case $C_1$ is $3.41 \times 10^{-3}$ and $3.32 \times 10^{-3}$, respectively. 
While, the ensemble heat flux $\langle q_w \rangle$ of $G_1$ and $C_1$ is $-0.0309$ and $-0.0396$, respectively. 
It is concluded that the thermal non-equilibrium three-temperature effects of $CO_2$ enlarge the ensemble heat transfer by $20\%$, and slightly decrease the ensemble frictional force.

For the first-order ensemble statistical quantities in supersonic three-temperature $CO_2$ turbulent channel flow, the profiles of key ensemble quantities are presented in Figure \ref{co2_cchannel_rhotma_rms}. 
The equilibrium temperature $T^{eq}$ is defined as Eq.\eqref{Temperature_eq}.
Compared with the one-temperature case $G_1$, we observe the large discrepancies in the near-wall density profiles (approximate $||Y/H|| \ge 0.8$) and the equilibrium temperature profiles in the off-wall region (approximate $||Y/H|| \le 0.95$).
Both the ensemble wall density and central equilibrium temperature decrease by approximate $15\%$.
It can be concluded that the thermal non-equilibrium three-temperature effects of supersonic $CO_2$ affect its ensemble thermal quantities greatly.
In Figure \ref{co2_cchannel_rhotma_rms}, the ensemble streamwise velocity in the log-law region of case $C_1$ is lower than that of case $G_1$, indicating that the ${\left\langle U \right\rangle}^+$ in log-law region is suppressed by the non-equilibrium three-temperature effects.
There is little change in the ensemble Mach number ${\left\langle Ma \right\rangle}$ and ensemble VD transformation of streamwise velocity ${\left\langle U \right\rangle}_{VD}^+$, showing that the VD transformation \cite{huang1994van} still works well for supersonic $CO_2$ turbulent channel flow.
Figure \ref{co2_cchannel_uplusvdrms} shows that the peak normalized r.m.s. of density $\rho_{rms}^{'}/ \langle \rho \rangle$ and equilibrium temperature ${(T^{eq})}^{'}_{rms}/ \langle T^{eq} \rangle$ of supersonic three-temperature $CO_2$ are suppressed near the wall region, and the corresponding peak locations are much closer to the wall than these of case $G_1$.
While, the peak of r.m.s. of Mach number $Ma_{rms}^{'}$ in case $C_1$ is larger than that of case $G_1$.
Compared with one-temperature supersonic turbulent channel flow, Figure \ref{co2_cchannel_uplusvdrms} shows that the normalized turbulent intensities of three-temperature $CO_2$ are suppressed above the $Y^+  \approx 20$, $Y^+  \approx 100$, and $Y^+  \approx 40$ regions for $U_{rms}^{'}/\langle u_{\tau} \rangle$, $V_{rms}^{'}/\langle u_{\tau} \rangle$, and $W_{rms}^{'}/\langle u_{\tau} \rangle$, respectively.
Correspondingly, the normalized Reynolds stress is suppressed above the $Y^+  \approx 120$ region.
In supersonic turbulent channel flow, it can be concluded that the thermal non-equilibrium three-temperature effects of $CO_2$ suppress the peak of normalized r.m.s. of density and temperature, normalized turbulent intensities and Reynolds stress.

\begin{figure}[!htp]
	\centering
	\includegraphics[width=0.485\textwidth]{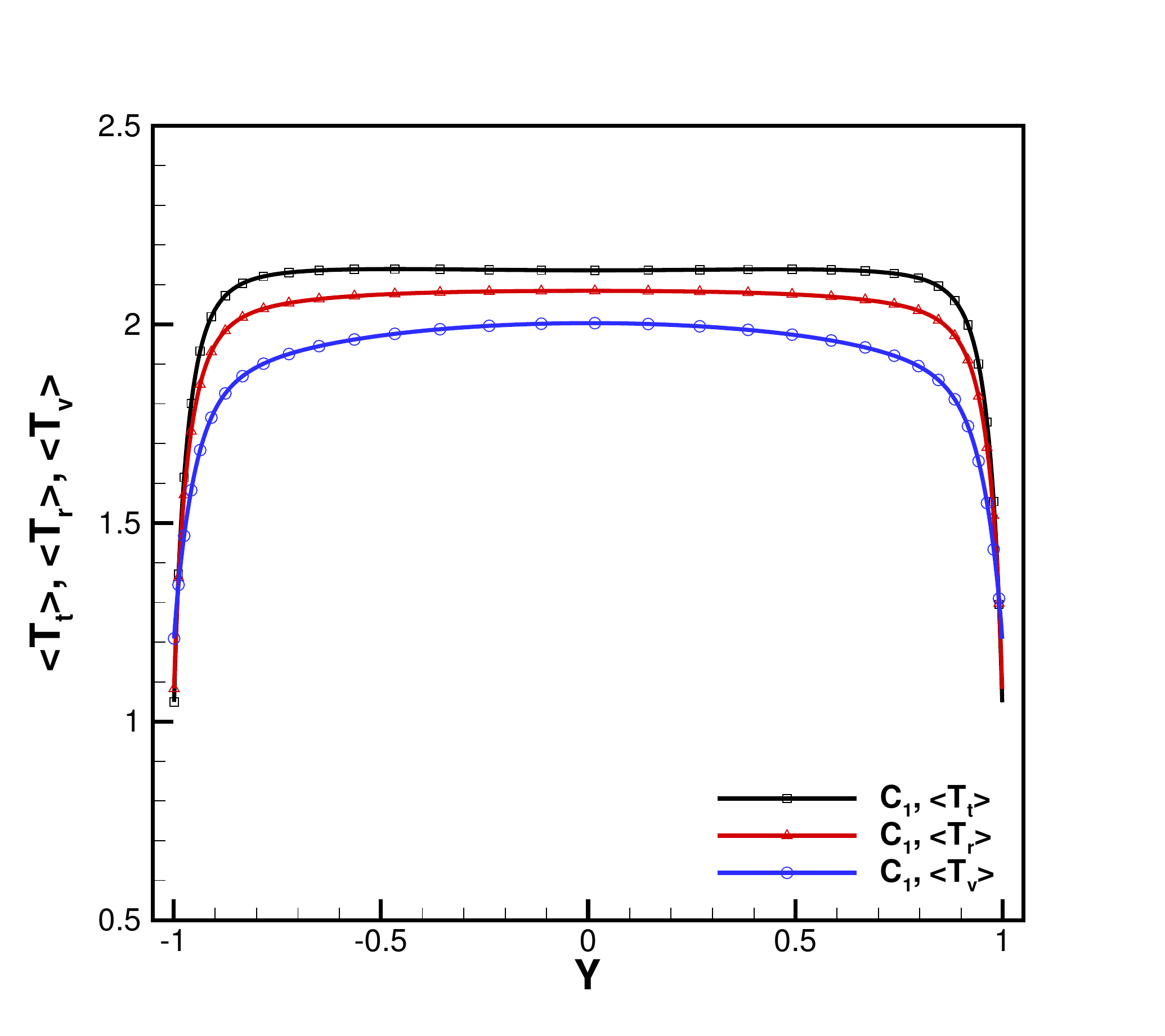}
	\includegraphics[width=0.485\textwidth]{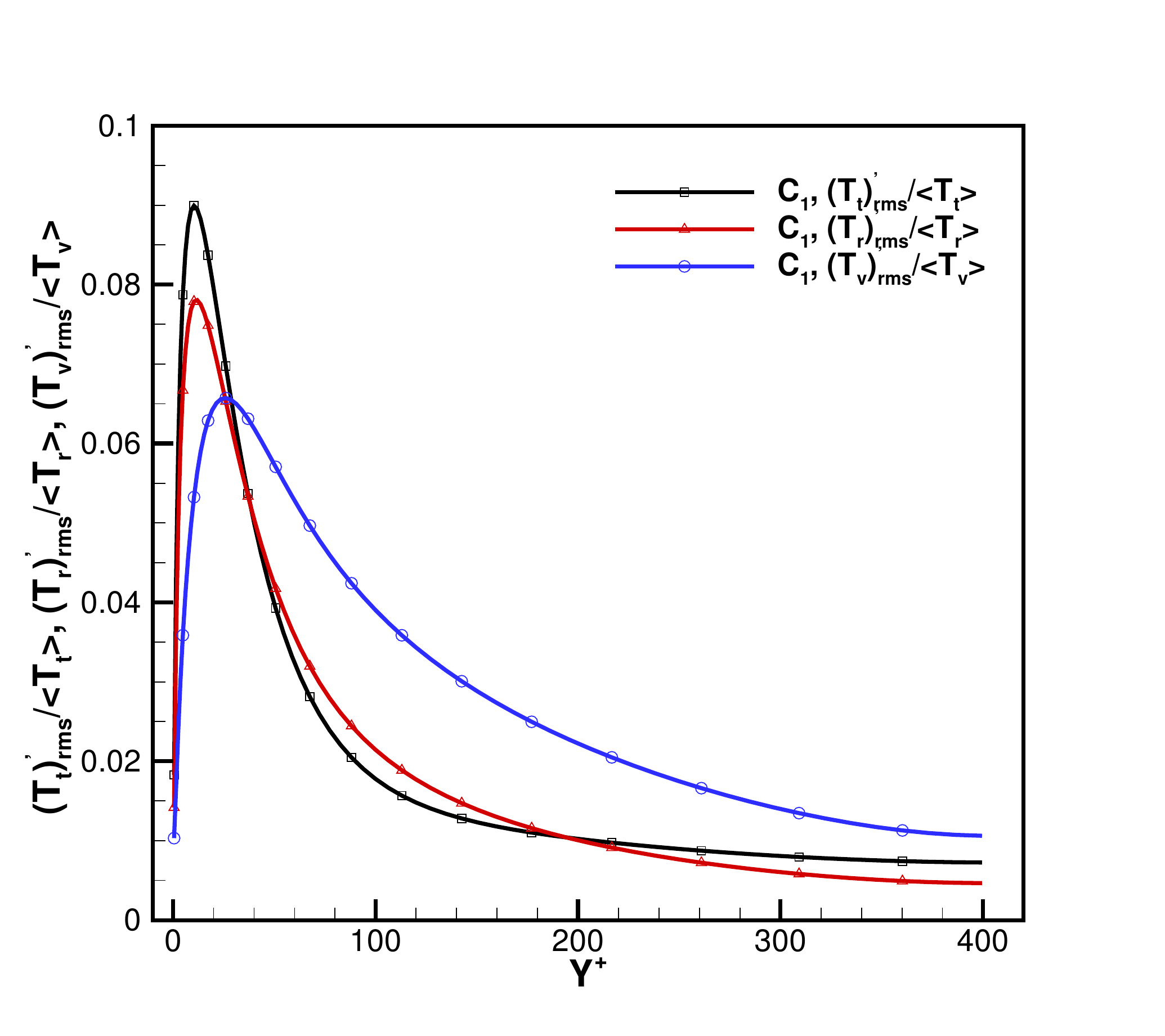}
	\vspace{-4mm}
	\caption{\label{co2_cchannel_Ttrb_rms} Supersonic thermal non-equilibrium three-temperature $CO_2$  turbulent channel flow: the ensemble translational, rotational, and vibrational temperature $\langle T_t \rangle$, $\langle T_r \rangle$, $\langle T_v \rangle$ (left), and the normalized r.m.s. of translational, rotational, and vibrational temperature $(T_t)^{'}_{rms}/ \langle T_t \rangle$, $(T_r)^{'}_{rms}/ \langle T_r \rangle$, $(T_v)^{'}_{rms}/ \langle T_v \rangle$ (right).}
\end{figure}
\begin{figure}[!h]
	\centering
	\includegraphics[width=0.49\textwidth]{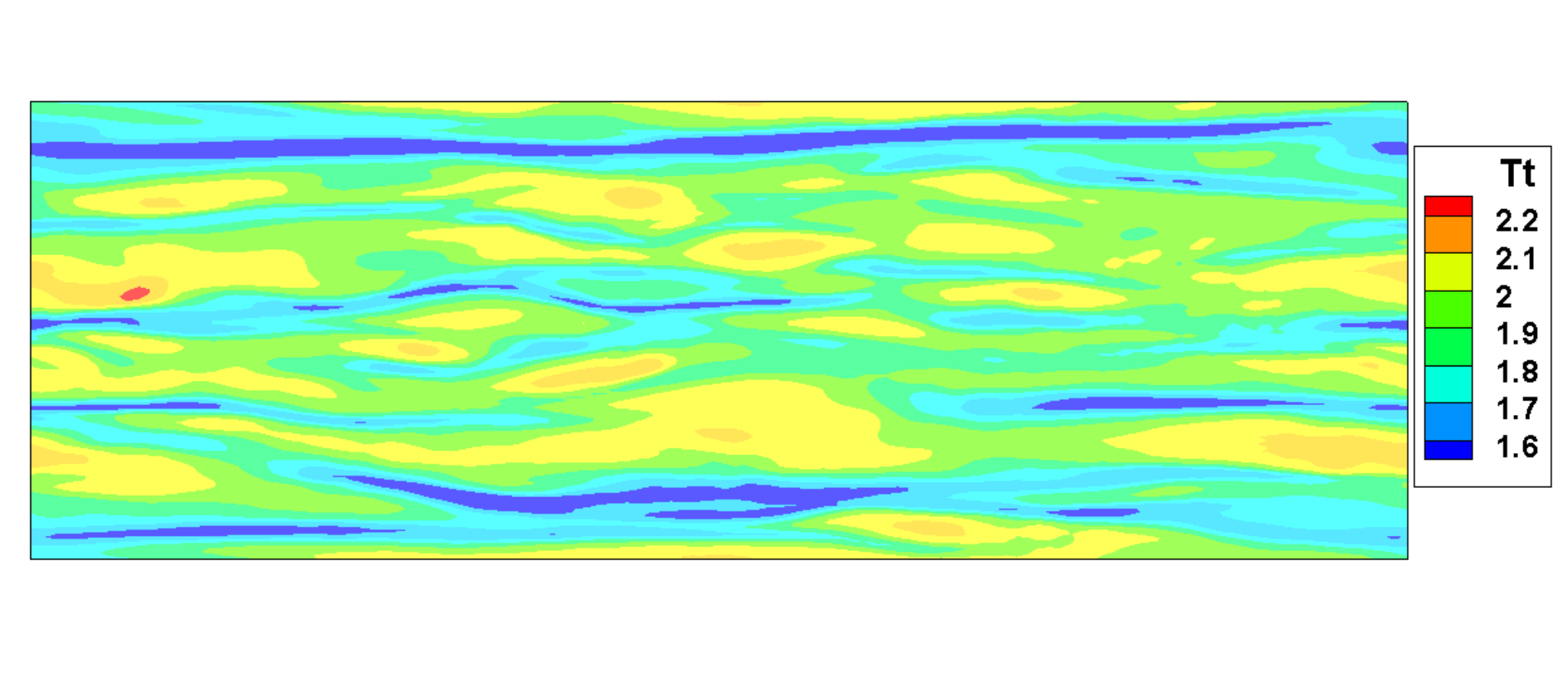}
	\includegraphics[width=0.49\textwidth]{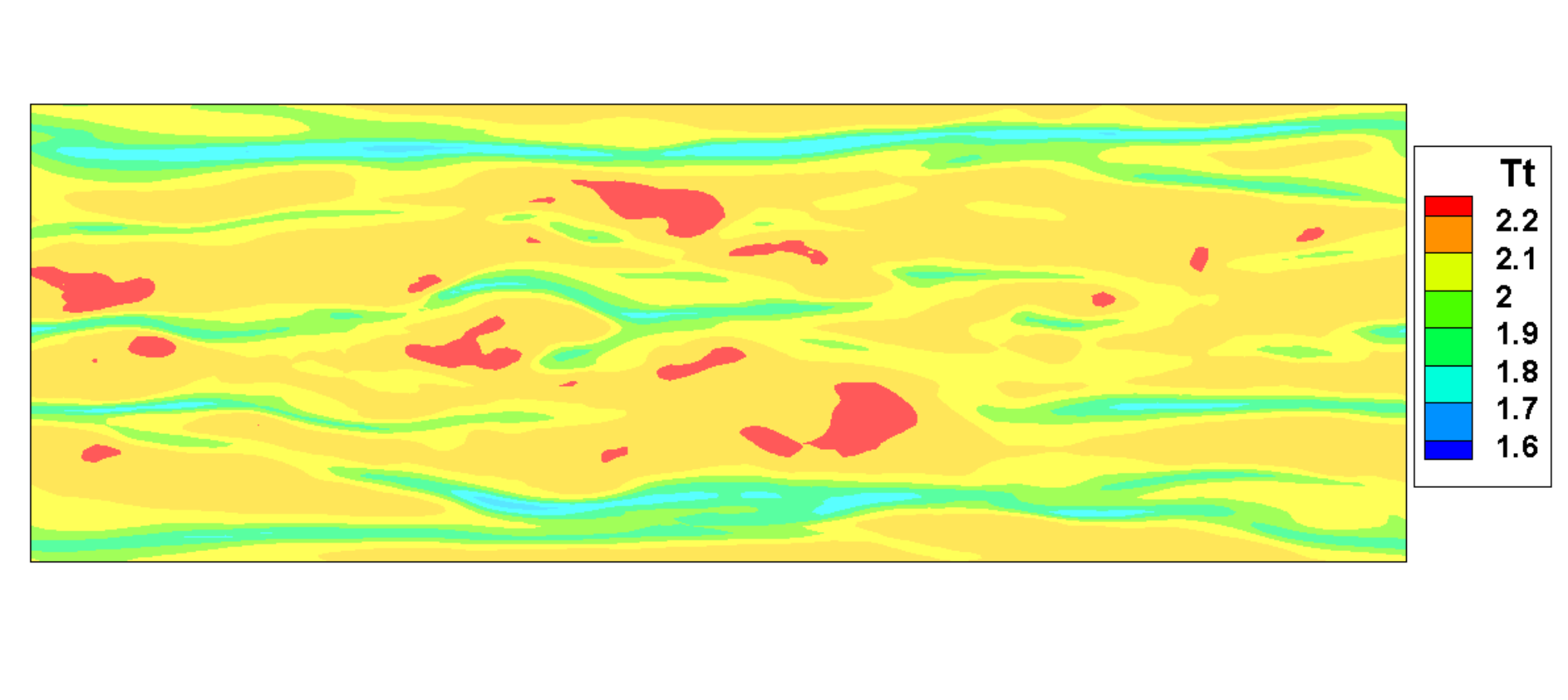}
	\includegraphics[width=0.49\textwidth]{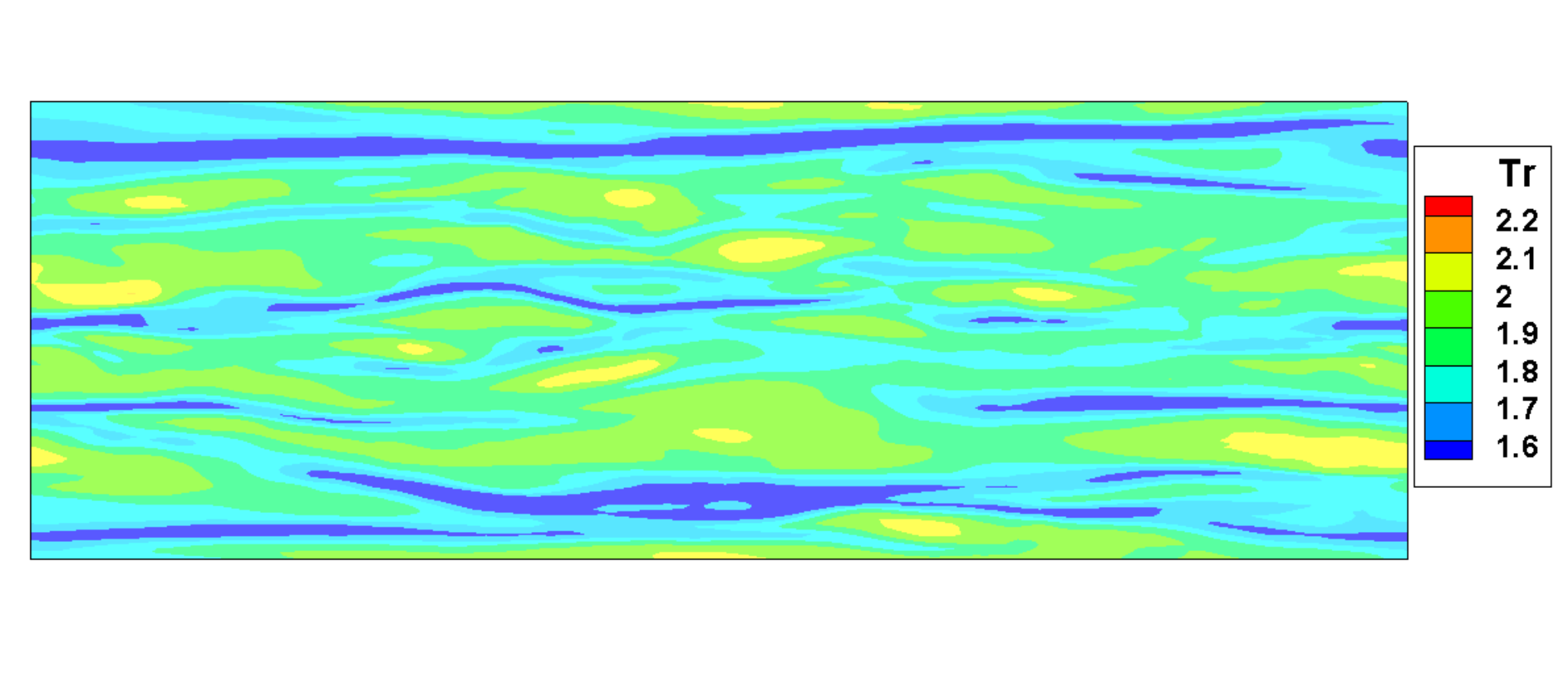}
	\includegraphics[width=0.49\textwidth]{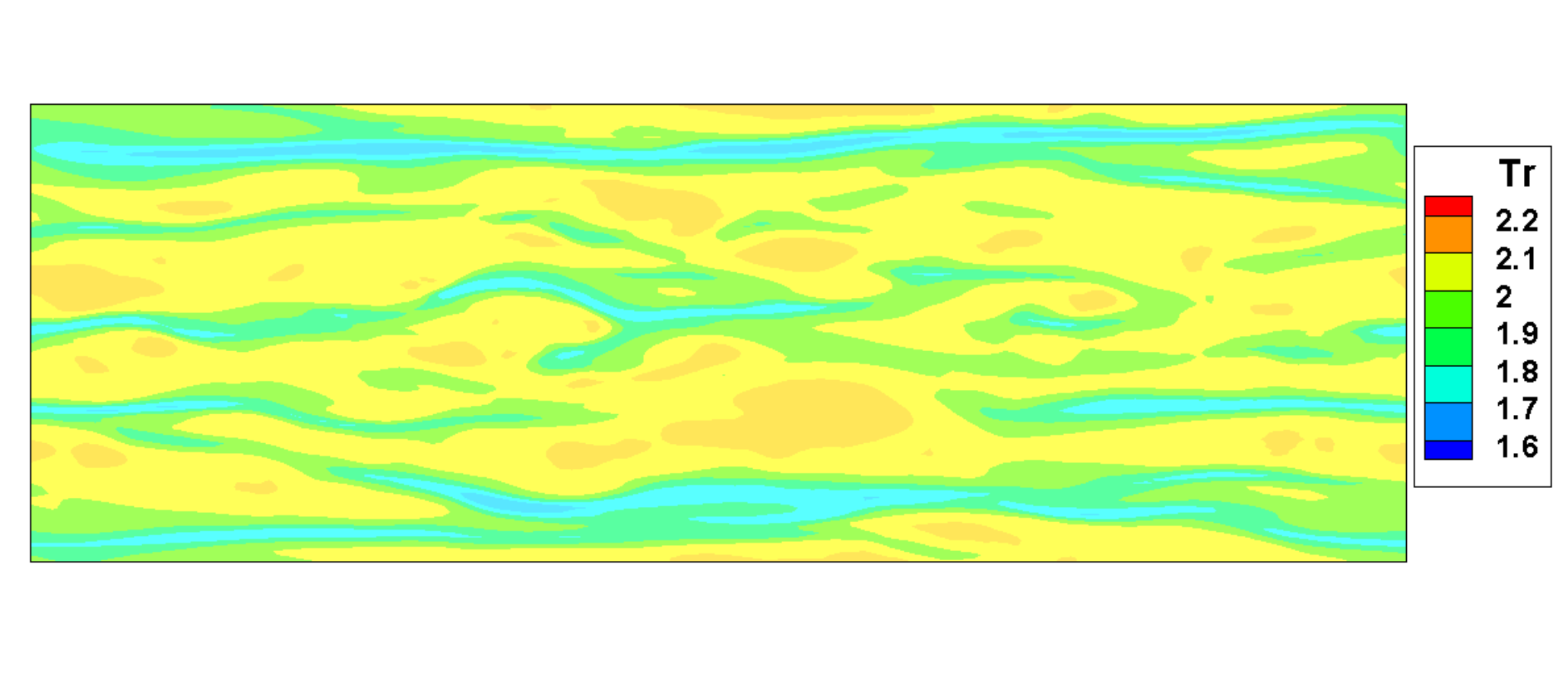}
	\includegraphics[width=0.49\textwidth]{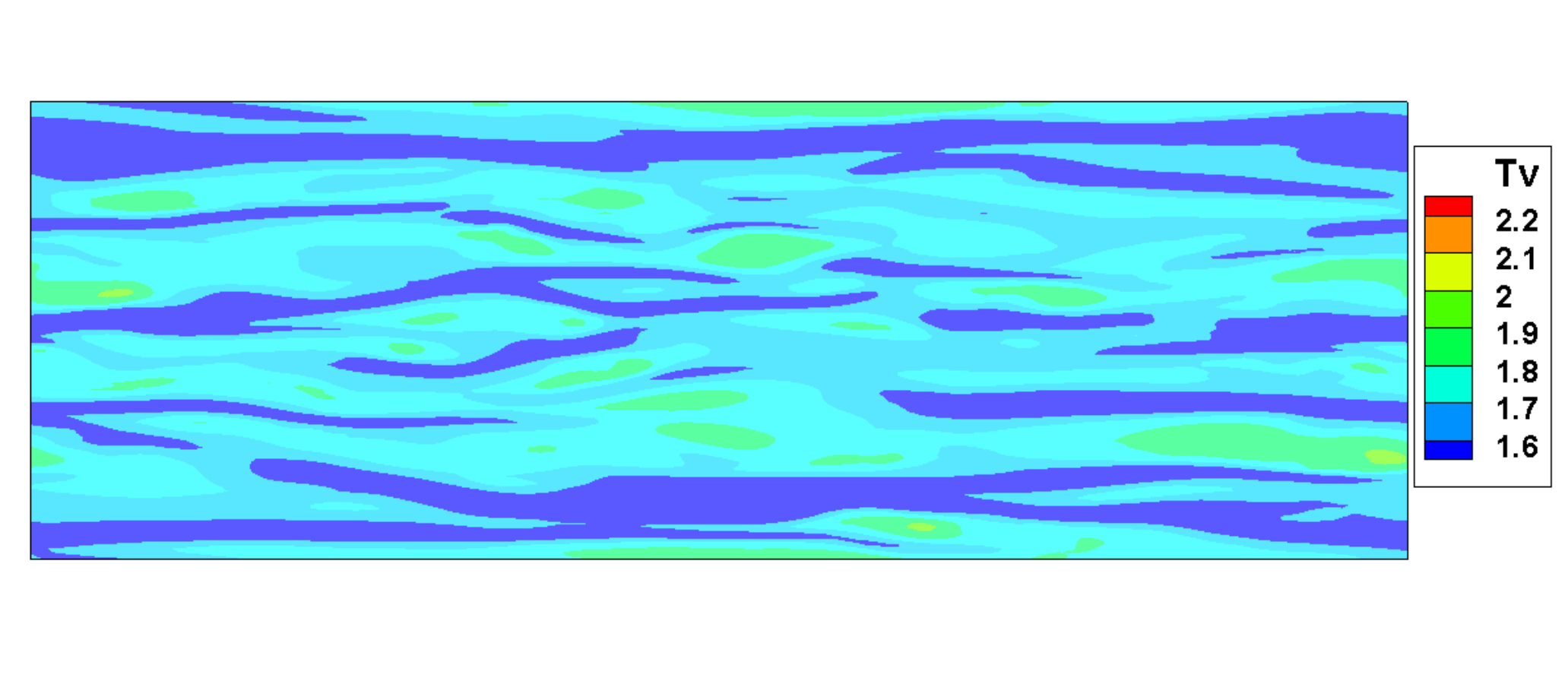}
	\includegraphics[width=0.49\textwidth]{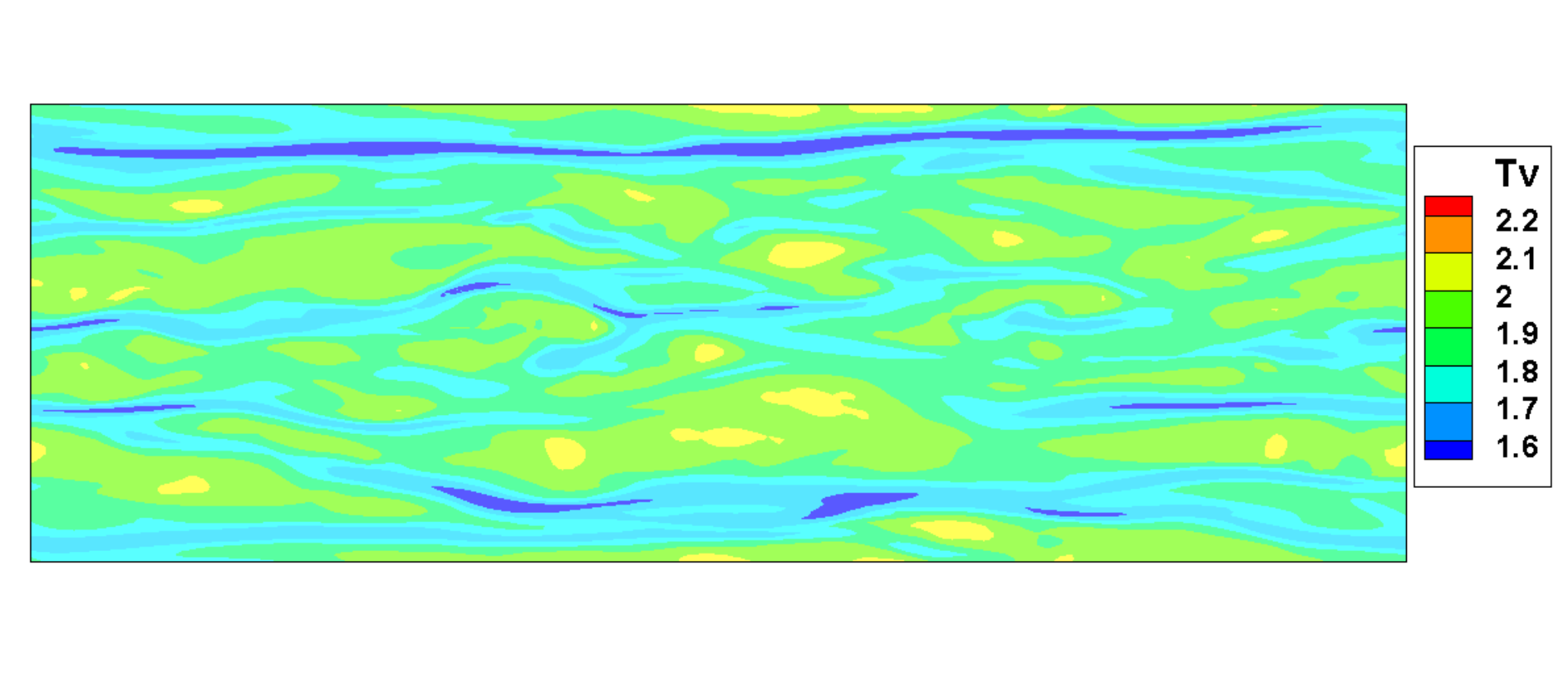}
	\vspace{-4mm}
	\caption{\label{co2_contours_2D3T-095} Supersonic thermal non-equilibrium three-temperature $CO_2$ turbulent channel flow: slices of translational, rotational, and vibrational temperatures at $Y^+ = 20.3$ (left column) and $Y^+ = 40.6$ (right column).}
\end{figure}
Be of special interest in the non-equilibrium performance of three internal energy modes of supersonic three-temperature $CO_2$ turbulent channel flow.
Figure \ref{co2_cchannel_Ttrb_rms} shows the ensemble average of translational, rotational and vibrational temperatures.
At the wall, the three ensemble temperatures sort from the high to low as $\langle T_v \rangle$, $\langle T_r \rangle$, $\langle T_t \rangle$, and the opposite order of three temperatures is observed in the off-wall region (approximate $||Y/H|| \le 0.95$).
Figure \ref{co2_cchannel_Ttrb_rms} as well shows that the peak of normalized r.m.s. of translational temperature $(T_t)^{'}_{rms} / \langle T_t \rangle$ and rotational temperature $(T_r)^{'}_{rms} / \langle T_r \rangle$ is almost co-located at $Y^+ \approx 15$.
While the peak location of normalized r.m.s. of vibrational temperature $(T_v)^{'}_{rms} / \langle T_v \rangle$ is far from the wall at $Y^+ \approx 30$.
The peak value of normalized r.m.s. of temperatures sorting from the high to low is  translational temperature, rotational temperature, and vibrational temperature.
Additionally, it is observed that $(T_v)^{'}_{rms} / \langle T_v \rangle$ is larger than $(T_t)^{'}_{rms} / \langle T_t \rangle$  and   $(T_r)^{'}_{rms} / \langle T_r \rangle$ above the $Y^+ \approx 30$ region.
Figure \ref{co2_contours_2D3T-095} clearly shows the streamwise low-temperature and high-temperature ribbon-like regions for all three temperatures.
At $Y^+ = 20.3$, the low-temperature ribbon-like regions are dominated in vibrational temperature fields, while the high-temperature ones dominate the translational temperature fields at $Y^+ = 40.6$.
Compared with the vibrational temperature fields,
Figure \ref{co2_contours_2D3T-095} obviously shows the rotational temperature fields have the higher similarity with translational temperature fields in temperature amplitude.
As shown in Figure \ref{fig_tau_trv_eta}, for $CO_2$, the rotational relaxation time $\tau_r$ is much closer with the translational relaxation time $\tau_t$.
However, the vibrational relaxation time $\tau_v$ is thousands of times larger than that of $\tau_t$.
Thus, the $T_t$ and $T_r$ are more likely to be equilibrium through the interaction between translational and rotational modes with closer relaxation time.
Meanwhile, the much longer relaxation process of vibrational modes may account for its small correlation with translational modes and rotational modes.

\section{Conclusion and remarks}
The present paper focuses on thermal non-equilibrium three-temperature effects of carbon dioxide ($CO_2$) in supersonic turbulent channel flow.
Essential ingredient has been addressed for compressible $CO_2$ turbulent flows, namely, the thermal non-equilibrium interactions among translational, rotational, and vibrational modes.
The three vibrational modes of $CO_2$ are addressed, and the double degenerated bending modes $\nu_2$ is equipped with the characteristic vibrational temperature $\theta_{v} = 959.66K$, which is much lower than that of $N_2$ and $O_2$.
Thus, the excitation of vibrational modes of $CO_2$ requires to be modeled and simulated carefully even under the room temperature.
The translational, rotational and vibrational relaxation time of $CO_2$ are calibrated.
Then, $CO_2$ is modeled in an extended three-temperature BGK-type model within the well-established kinetic framework.
To achieve high-order accuracy in space and time for simulating $CO_2$ turbulence, the non-equilibrium high-accuracy GKS has been constructed with the second-order kinetic flux, fifth-order WENO-Z reconstruction, and two-stage fourth-order time discretization.

With implementing the non-equilibrium high-accuracy GKS in the large-scale parallel in-house solver, the DNS of supersonic $CO_2$ turbulent channel flow is conducted.
Compared with the one-temperature supersonic turbulent channel flow, the three-temperature effects of $CO_2$ are analyzed.
The ensemble frictional force and ensemble heat flux of the wall, as well as the typical ensemble and fluctuating turbulent quantities of supersonic $CO_2$ turbulent channel flow are investigated.
Thermal non-equilibrium three-temperature effects of $CO_2$ enlarges the ensemble heat transfer of the wall by approximate $20\%$, and slightly decreases the ensemble frictional force.
The ensemble density and temperature fields are greatly affected, and both the ensemble wall density and central equilibrium temperature decrease by approximate $15\%$.
The ensemble streamwise velocity in log-law region is suppressed.
There is little change in VD transformation of streamwise velocity, which shows that the VD transformation still works well for supersonic $CO_2$ turbulent channel flow.
We observe that the peak of normalized r.m.s. of density and temperature, normalized turbulent intensities and Reynolds stress are suppressed in supersonic three-temperature $CO_2$ turbulent flow.

Numerical simulation confirms the thermal non-equilibrium three-temperature performance of $CO_2$.
The streamwise low-temperature and high-temperature ribbon-like regions are clearly observed for all three temperatures near the wall region.
The vibrational modes of $CO_2$ behave quite differently with rotational and translational modes.
The peak positions of normalized r.m.s. of translational temperature and rotational temperature are almost co-located.
Both the maximum ensemble temperature and normalized r.m.s. temperature sort from high to low is translational temperature, rotational temperature, and vibrational temperature. 
Compared with the vibrational temperature fields, the rotational temperature fields have the higher similarity with translational temperature fields both in temperature amplitude and its structure.
The much longer relaxation process of vibrational modes of $CO_2$ may account for its small correlation with translational modes and rotational modes.

In the future, we expect to explore the interdisciplinary studies on $CO_2$ transition in supersonic/hypersonic flat plate, i.e., the overheating phenomenon at the late transitional period \cite{franko2013breakdown}.
The multi-scale numerical framework UGKS and UGKWP \cite{xu2010unified, chen2020three} also provide the solid foundation in further multi-scale $CO_2$ flows, i.e., the Mars re-entry vehicles from rarefied to continuum regimes.
In addition, the realistic wall boundary conditions for rotational and vibrational internal energy of $CO_2$ deserve to be explored for more delicate simulations in wall-bounded $CO_2$ turbulence.

\section*{Ackonwledgement}
Many thanks to Dr. Zhao Wang for helpful discussions on the wall boundary conditions of $CO_2$ gas flows.
This research is supported by the National Numerical Windtunnel project, the Department of Science and Technology of Guangdong Province (2020B1212030001).
The authors would like to acknowledge TaiYi supercomputers in the SUSTech for providing high performance computational resources.

\section*{Appendix A. Sutherland law and curve-fit shear viscosity of $CO_2$}\label{co2_appendix_shther}
For dilute $CO_2$ gas flows, Sutherland law \cite{white2006viscous} gives the shear viscosity  as
\begin{equation}\label{shear_co2_1}
	\mu(T_t) = \mu_0 (\frac{T_t}{T_0})^{\frac{3}{2}} \frac{T_0 + S}{T_t + S},
\end{equation}
where the reference shear viscosity $\mu_0 = 1.370 \times 10^{-5} kg/(m \cdot s)$, $T_0 = 273K$ and $S = 222K$ is approximately valid between $190 K$ and $1700 K$.
In terms of air, it is noted that $\mu_0 = 1.716 \times 10^{-5} kg/m \cdot s$, $T_0 = 273K$ and $S = 111K$ for air is approximately valid between $210 K$ and $1900 K$.
Sutherland law gives the thermal conductivity \cite{white2006viscous} as
\begin{equation}\label{thercon_co2_1}
	\kappa(T_t) = \kappa_0 (\frac{T_t}{T_0})^{\frac{3}{2}} \frac{T_0 + S}{T_t + S},
\end{equation}
where the reference thermal conductivity $\kappa_0 = 1.46 \times 10^{-2} W/(m \cdot K)$, $T_0 = 273K$ and $S = 1800K$ is approximately valid between $180 K$ and $700 K$.
In terms of air, it is noted that $\kappa_0 = 2.41 \times 10^{-2} W/(m \cdot K)$, $T_0 = 273K$ and $S = 194$ is approximately valid between $160 K$ and $2000 K$.

It is reported that the shear viscosity of $CO_2$ \cite{candler1990computation} is well approximated over a wide range of temperature up to $20000K$.
The curve-fit expression of shear viscosity reads
\begin{equation}\label{shear_co2_3}
	\mu(T_t) = 0.1 exp\{(A ln T_t + B) ln T_t + C\},
\end{equation}
where $A = -0.01952739$, $B = 1.047818$, and $C = -14.32212$.
In Eq.\eqref{shear_co2_3}, shear viscosity $\mu$ is in the unit $kg/(m \cdot s)$.

\section*{Appendix B. Calibrated rotational relaxation time of $CO_2$}\label{co2_appendix_rot}
This appendix provides the power law for rotational relaxation time $\tau_r$ of $CO_2$.
For bulk viscosity $\eta_b^r$ arising from the rotational modes, the calibration data of $CO_2$ \cite{wang2019bulk} is utilized as Table \ref{co2_tau_r_calibration}.
Based on the least-square method, with $T_0 = 273K$, calibration data in Table \ref{co2_tau_r_calibration} provides the $\tau_{r0}= 2.99 \times 10^{-10} s$ and $n_r = 4.47$.
It should be noted that the data of bulk viscosity in high temperature is not adequate.
\begin{table}[!h]
	\centering
	\begin{tabular}{c|c|c|c|c|c|c}
		\hline \hline
		$T_t/K$    &$258.05$     &$274.36$  &$293.24$   &$312.80$  &$330.73$ &$353.15$\\
		\hline
		$\eta_b^r/\mu$    &$0.226$          &$0.180$   &$0.188$  &$0.198$ &$0.200$  &$0.191$\\
		\hline \hline
	\end{tabular}
	\caption{\label{co2_tau_r_calibration} Calibration data \cite{wang2019bulk} for determining the rotational relaxation time $\tau_r$ of $CO_2$.}
\end{table}

To absorb the high-temperature information of $\tau_r$, the power $n_r$ should be corrected based on the refereed calculation of $CO_2$ \cite{kustova2021investigation}.
The rotational relaxation time is used as $\tau_r = Z_r \tau_t$, where the rotational collision number $Z_r$ is given by Parker equation \cite{parker1959rotational} as
\begin{align}
	Z_r = \frac{Z_r^{\infty}}{1 + (\pi^{3/2}/2) \sqrt{T^{\ast}/T_t} + (\pi + \pi^2/4)(T^{\ast}/T_t)}.
\end{align}
$T^{\ast}$ is the characteristic temperature of inter-molecular potential, and $Z_r^{\infty}$ is the limiting value. 
The values $Z_r^{\infty} = 20.39 $ and $T^{\ast} = 91.5K$ are used for $CO_2$ \cite{kustova2021investigation}.
Thence, with the fixed $T_0 = 273K$ and $\tau_{r0} = 2.99 \times 10^{-10} s$, the $Zr \tau_t = \tau_{r0}(T/T_0)^{n_r}$ gives the corrected power $n_r = 1.59$ at $T = 2000K$.

\section*{Appendix C. Connection between macroscopic variables and mesoscopic coefficients}\label{co2_appendix_mesoscopic_coefficients}
The connection between the spatial derivatives of macroscopic flow variables and the expansion of intermediate equilibrium distribution function $f^v$ as Eq.\eqref{intermediate_f_v} reads
\begin{equation}\label{micro_coefficients} 
	\frac{\partial \boldsymbol{Q}}{\partial x_i}
	=\int \frac{\partial f^v}{\partial x_i}\boldsymbol{\psi_v}\text{d}\Xi_v
	\equiv \int a f^v\boldsymbol{\psi_v}\text{d}\Xi_v,
\end{equation}
where $a$ denotes the spatial mesoscopic coefficients in Eq.\eqref{formalsolution_neq} as
\begin{align}
	a=\boldsymbol{a}^T \boldsymbol{\tilde{\psi}_v}=a_1+ a_2 u_1 + a_3 u_2 + a_4 u_3 + a_5 (u_1^2 + u_2^2 + u_3^2) + a_6 \xi_r^2 + a_7 \xi_v^2.
\end{align}
Eq.\eqref{micro_coefficients} can be rewritten into following linear system 
\begin{equation}\label{micro_coefficients_matrix} 
	\frac{1}{\rho}\frac{\partial \boldsymbol{Q}}{\partial x_i} =\Big(\frac{1}{\rho}\int\boldsymbol{\psi_v} \otimes \boldsymbol{\tilde{\psi}_v}^T f^v \text{d}\Xi_v \Big)\boldsymbol{a} 
	\equiv \boldsymbol{M} \boldsymbol{a},
\end{equation}
Each component of $(a_1, …, a_6, a_7)^T$ in Eq.\eqref{micro_coefficients_matrix} can be determined uniquely 
\begin{equation}
	\left\{
	\begin{aligned}
		a_7 &= \frac{4 \lambda_v^2}{N_v} B_6, \\
		a_6 &= \frac{4 \lambda_r^2}{N_r} B_5, \\
		a_5 &= \frac{4 \lambda_t^2}{3} (B_4 -  U_1 B_1 - U_2 B_2 - U_3 B_3 - B_5 - B_6), \\
		a_4 &= 2 \lambda_t B_3 - 2 U_3 a_5, \\
		a_3 &= 2 \lambda_t B_2 - 2 U_2 a_5, \\
		a_2 &= 2 \lambda_t B_1 - 2 U_1 a_5, \\
		a_1 &= \frac{1}{\rho}\frac{\partial \rho}{\partial x_i} - U_1 a_2 - U_2 a_3 - U_3 a_4 - (U_1^2 + U_2^2 + U_3^2 + \frac{3}{2 \lambda_t}) a_5 - \frac{N_r}{2 \lambda_r} a_6 - \frac{N_v}{2 \lambda_v} a_7,
	\end{aligned}
	\right.
\end{equation}
with
\begin{equation}
	\left\{
	\begin{aligned}
		B_1 &= \frac{1}{\rho} [\frac{\partial (\rho U_1)}{\partial x_i} - U_1 \frac{\partial \rho}{\partial x_i}], \\
		B_2 &= \frac{1}{\rho} [\frac{\partial (\rho U_2)}{\partial x_i} - U_2 \frac{\partial \rho}{\partial x_i}], \\
		B_3 &= \frac{1}{\rho} [\frac{\partial (\rho U_3)}{\partial x_i} - U_3 \frac{\partial \rho}{\partial x_i}], \\
		B_4 &= \frac{1}{\rho} [\frac{\partial (\rho E)}{\partial x_i} - E \frac{\partial \rho}{\partial x_i}], \\
		B_5 &= \frac{1}{\rho} [\frac{\partial (\rho E_r)}{\partial x_i} - E_r \frac{\partial \rho}{\partial x_i}], \\
		B_6 &= \frac{1}{\rho} [\frac{\partial (\rho E_v)}{\partial x_i} - E_v \frac{\partial \rho}{\partial x_i}].
	\end{aligned}
	\right.
\end{equation}

For the temporal mesoscopic coefficient in Eq.\eqref{formalsolution_neq}, the relation between temporal derivatives of macroscopic variables and $f^v$ can be written as
\begin{equation}\label{micro_coefficients_temp} 
	\frac{\partial \boldsymbol{Q}}{\partial t}=\int \frac{\partial f^v}{\partial t}\boldsymbol{\psi_v}\text{d}\Xi_v
	\equiv \int A f^v\boldsymbol{\psi_v}\text{d}\Xi_v,
\end{equation}
with
\begin{align}
	A = \boldsymbol{A}^T \boldsymbol {\tilde{\psi}_v}=A_1+ A_2 u_1 + A_3 u_2 + A_4 u_3 + A_5 (u_1^2 + u_2^2 + u_3^2) + A_6 \xi_r^2 + A_7 \xi_v^2.
\end{align}
The temporal derivatives of macroscopic variables can be given according to the compatibility condition as
\begin{equation}
	\begin{aligned}
		\int (\frac{\partial{f^v}}{\partial t} + u_i \frac{\partial f^v}{\partial x_i}) \boldsymbol{\psi_v} \text{d}\Xi_v = \boldsymbol{0}.
	\end{aligned}
\end{equation}
In a similar way, the above components $(A_1, …, A_6, A_7)^T$ in Eq.\eqref{micro_coefficients_temp} can be determined uniquely.

\bibliographystyle{unsrt}
\bibliography{caogybib}
\end{document}